\documentclass[usenatbib]{mn2e}
\usepackage{graphicx,times}
\usepackage{bm}
\usepackage{epsf}

\newcommand{\re}{\mathrm {Real}}
\newcommand{\on}{{\hat \Omega}}
\newcommand{\beq}{\begin{equation}}
\newcommand{\eeq}{\end{equation}}
\newcommand{\beqa}{\begin{eqnarray}}
\newcommand{\eeqa}{\end{eqnarray}}
\newcommand{\n}{\noindent}
\newcommand{\nn}{\nonumber}
\newcommand{\oh}{\hat \Omega}

\def\qtwo{\qquad\qquad}

\def\be{\begin{equation}}
\def\ee{\end{equation}}

\def\ben{\begin{eqnarray}}
\def\een{\end{eqnarray}}

\def\myK{{\cal K}}
\def\myC{{ \cal C}}

\def\ho{{\hat \Omega}}

\def\bk{{\bf k}}
\def\bK{{\bf K}}

\def\2p{{(2\pi)^2}}

\def\la{{\langle}}
\def\ra{{\rangle}}

\date{\today,~ $ $Revision: 0.9 $ $}

\begin{document}

\onecolumn

\title[New Optimised Estimators for the Trispectrum: Beyond Lowest Order of Non-Gaussianity]
{New Optimised Estimators for the Primordial Trispectrum}

\author[Munshi et al.]
{Dipak Munshi$^{1,2}$, Alan Heavens$^{1}$, Asantha Cooray$^{3}$, Joseph Smidt$^{3}$, Peter Coles$^{1}$, Paolo Serra$^3$ \\
$^{1}$ Scottish Universities Physics Alliance (SUPA),~ Institute for Astronomy, University of Edinburgh, 
Blackford Hill,  Edinburgh EH9 3HJ, UK \\
$^{2}$ School of Physics and Astronomy, Cardiff University, CF24 3AA \\
$^{3}$ Department of Physics and Astronomy, University of California, Irvine, CA 92697}
\maketitle

\begin{abstract}
Cosmic microwave background studies of non-Gaussianity involving higher-order multispectra can distinguish between
 early universe theories that predict nearly identical power spectra. However, the recovery of higher-order multispectra is difficult from realistic data due to
their complex response to inhomogeneous noise and partial
sky coverage, which are often difficult to model analytically. A traditional alternative is to use
one-point cumulants of various orders, which collapse the information present in a multispectrum
to one number.  The disadvantage of such a radical compression of the data is a loss of information as to the source of the statistical behaviour. A recent study by Munshi \& Heavens (2009)
has shown how to define the skew spectrum (the power spectra of a certain cubic field, related to the bispectrum)
in an optimal way and how to estimate it from realistic data. The skew spectrum retains some of the information from the
full configuration-dependence of the bispectrum, and can contain all the information on non-Gaussianity. In the present study, we extend the results
of the skew spectrum to the case of two degenerate power-spectra related to the trispectrum.
We also explore the relationship of these power-spectra and cumulant correlators previously used to study non-Gaussianity
in projected galaxy surveys or weak lensing surveys. We construct nearly
optimal estimators for quick tests  and generalise them to estimators which can handle realistic
data with all their complexity in a completely optimal manner. Possible generalisations for
arbitrary order are also discussed. We show how these higher-order statistics and the related
power spectra are related to the Taylor expansion coefficients of the potential in inflation models, and demonstrate how the trispectrum can constrain both the quadratic and cubic terms. 
\end{abstract}

\section{Introduction}

The inflationary paradigm which solves the flatness, horizon and monopole problem
makes clear predictions about the generation and nature of density perturbations \citep{Guth81,Staro79,Linde82,AlSt82,Sato81}.
Inflationary models predict the statistical nature of these
fluctuations, which are being tested against data from a range of recent
cosmological observations, including the recently-launched all-sky cosmic microwave background (CMB)
survey Planck\footnote {http://www.rssd.esa.int/index.php?project=Planck}. 
Various ground-based and space-based observations have already 
confirmed the generic predictions of inflation, including a flat or nearly flat 
universe with nearly scale-invariant adiabatic perturbations at large angular scales.
Several planned missions are also targeting the detection of the gravitational
wave background through polarisation experiments - another generic prediction
of inflationary models. The other major prediction of the inflationary scenarios 
is the nearly Gaussian nature of these perturbations. In the standard
slow-roll paradigm, the scalar field responsible for inflation fluctuates
with a minimal amount of self interaction which ensures that any non-Gaussianity
generated during the inflation through self-interaction would be small
\citep{Salopek90,Salopek91,Falk93,Gangui94,Acq03,Mal03}. See \cite{Bartolo06} for a recent
review and more detailed discussion. Any detection of non-Gaussianity would therefore 
be a measure of self-interaction or non-linearities involved, which can come from 
various alternative scenarios such as curvaton mechanism, warm inflation, ghost inflation as well as string theory
inspired D-cceleration and Dirac Born Infield (DBI) models of inflation
\citep{lindemukha,GuBeHea02,Lyth03}.

Early observational work on detection of primordial non-Gaussianity,
from COBE \citep{Komatsu02} and MAXIMA \citep{Santos} was followed by much more 
accurate analysis with WMAP\footnote{http://map.gsfc.nasa.gov/}\citep{Komatsu03,Crem07a,Spergel07}.
Optimised 3-point
estimators were introduced by \citet{Heav98}, and have been successively developed
\citep{KSW,Crem06,Crem07b,SmZaDo00,SmZa06}. Indeed, now an estimator for $f_{NL}$ which saturates
the Cramer-Rao bound has been found, capable of treating partial sky coverage and inhomogeneous noise \citep{SmSeZa09}.
The recent claim of a detection of non-Gaussianity in WMAP data \citep{YaWa08} has given a tremendous 
boost to the study of  primordial non-Gaussianity, as it can lift the degeneracy between various early-universe
theories which predict nearly the same primordial power spectrum. Most detection strategies focus on lowest order in
non-Gaussianity, i.e. the bispectrum or three-point correlation functions
\citep{KSW,Crem03,Crem06,MedeirosContaldi06,Cabella06,Liguori07,SmSeZa09}.
This is primarily because of a decrease in signal-to-noise as we move
up in the hierarchy of correlation functions - the higher-order correlation
functions are more dominated by noise than their lower-order counterparts.
Another related complication arises from the necessity to
optimise such estimators, and the impact of inhomogeneous noise and partial sky coverage is always
difficult to include in such estimates.

The recent study by \cite{MuHe09} suggested the possibility of
finding optimised cumulant correlators associated with higher-order multi-spectra
in the context of CMB studies. These correlators are well-studied in
the context of projected surveys such as projected galaxy surveys or
in the context of weak lensing studies using simulated maps \citep{Mu00,MuJe00,MuJe01}.
Early studies involving cumulant correlators focused mainly on 
understanding gravity-induced clustering in collisionless media and
were widely employed in many studies involving numerical simulations.
Cumulant correlators are multipoint correlation functions, collapsed to two points 
Although they are two-point correlators, they carry information
on the corresponding higher-order correlation functions. Due to
their {\em reduced} dimensionality they do not carry all the information that
is encoded in higher-order correlation function, but they carry
more than their one-point counterparts, namely the moments of the 
probability distribution function which are often used as clustering
statistics \citep{BernardReview}. 

One of the reasons to go beyond the lowest order in non-Gaussianity was pointed
out by many authors, including. e.g., \cite{RiqSper}. 
At smaller angular scales, the secondary effect may dominate \citep{SperGold99a,SperGold99b},
in direct contrast to larger angular scales where the anisotropies are generated
mainly at the surface of the last scattering. These secondary perturbations
are produced by interaction of CMB photons at much lower redshift with
the intervening large-scale matter distribution. Such effects will be directly
observable with Planck. The deflection of
CMB photons by the large-scale mass distribution offers the possibility of 
studying the statistics of density perturbations in an unbiased way and provide
clues to growth of structure formation for most part of the cosmic history.
Weak lensing of the CMB can provide valuable information for constraining
neutrino mass, dark energy equation of state and also
has the potential to assist detection of primordial gravitation waves through
CMB polarisation information; see e.g. \cite{LewChal} for a recent review. 
However weak lensing studies using the CMB 
need to address the contamination produced by other secondaries such as the
thermal Sunyaev Z\'eldovich (tSZ) effects and kinetic Sunayev Z\'eldovich
effect (kSZ) as well as by point sources, Although it is believed that
these contaminations are not so important in case of polarisation studies.
It was pointed out in \cite{RiqSper} that a real space statistic such as
$\langle \delta T^3(\oh) \delta T(\oh') \rangle_c$ (which is a cumulant correlator
of order four) can be used to separate the kSZ effect or Ostriker-Vishniac (OV) effect from the lensing effect
as the lensing contribution cancels out at the lowest order.
It was shown that, in addition to quantifying and controlling the kSZ contamination
of lensing statistics such statistics could as well play a very important role
in providing new insight into the history of reionization. In a completely
different context it was shown that this estimator also has use for 
testing models of primordial non-gaussianity using redshifted 21cm observations \citep{Coo21,Cooray8}. 
While cumulant correlators at third order 
$\langle \delta T^2(\oh) \delta T(\oh') \rangle_c$ can provide information regarding the non-Gaussianity parameter
$f_{\rm NL}$, fourth-order statistics such as $\langle \delta T^2(\oh) \delta T^2(\oh') \rangle_c$
can go beyond the lowest order non-Gaussianity by putting constraints on the
next-order parameter $g_{\rm NL}$ (to be introduced in later sections), albeit at lower signal-to-noise. However
with ongoing CMB missions such as Planck the situation will improve
and developing optimal methods for such higher-order cumulant correlators 
is the first step in this direction. 
There are now several studies which provide independent estimates of $f_{\rm NL}$
however we still lack such constraints for $g_{\rm NL}$.
This clearly is related to the fact that in typical models $g_{\rm NL}$ is 
expected to be small,  $g_{\rm NL} \le r /50 $, where $r$ is the scalar-to-tensor ratio \citep{Seery07}.
We also note that various studies have pointed out the link between the 
non-Gaussianity analysis and the estimators which test the anisotropy
of the primordial universe. We plan to address these issues in a related publication.
Several inflationary models provide direct consistency relations between $f_{\rm NL}$ and
$g_{\rm NL}$, e.g. $g_{NL} = (6f_{\rm NL}/5)^2 $ (in some publications it is also denoted as $f_2$ or $\tau_{NL}$ e.g. 
\citep{huOka02,Coo21,Seery07}). Testing of these consistency 
relations can give valuable clues to the mechanism behind the generation of initial perturbations.
However, a difficulty for methods designed to detect non-Gaussianity in the CMB is that other processes can contribute, 
such as gravitational lensing, unsubtracted point sources, and imperfect subtraction of 
galactic foreground emission  discussed by, e.g., \citet{GoldbergSpergel99,CoorayHu,VerdeSperg02,Castro04,Babich08}.

While the main motivation in this work is to study the primordial trispectrum, we note that
mode-mode coupling resulting from weak lensing of the CMB produces a trispectrum which 
has been studied using power-spectra associated with $\langle \delta T(\oh)^2 \delta T(\oh')^2 \rangle_c$.
Lensing studies involving the CMB can achieve higher signal-to-noise ratio at the 
level of the bispectrum if we use external data sets to act as a tracers of large-scale
structure.  However, the lowest order at which the internal detection of CMB lensing is possible 
is the trispectrum.   There have been some attempts to detect non-gaussianity at the level of the trispectrum using
e.g. COBE 4yr data release \citep{Kunz01}, BOOMERanG data \citep{Troia03} and more recently using WMAP 
3 year data \citep{Spergel07}.

The layout of this paper is as follows. The section \textsection2 we introduce the concept of
the higher-order cumulant correlators and how they are linked to corresponding correlation functions in real space. 
In \textsection3 we discuss the harmonic transforms
of the cumulant correlators and their relations to the corresponding multi-spectra. We also discuss estimators based on pseudo-$C_l$ 
(PCL) estimators used for power spectrum estimation and generalise them to two-point cumulant correlators
at higher order in this section. In \textsection4 we briefly discuss the ``{\em local}'' models for initial 
perturbations and the
resulting trispectrum. These models are then used in \textsection4 to optimise the power spectra
associated with the multi-spectra. This approach is nearly optimal, describing the mode-mode coupling using the
``{\em fraction of sky}'' approach familiar from other studies. In \textsection5
we present an estimator which is nearly optimal and can take into account partial sky coverage as well as
realistic inhomogeneous noise resulting from the scanning strategy. This estimator can work directly with any specific theoretical model for primordial
trispectra, based on concepts of matched filtering, and 
generalises the results obtained previously by \cite{MuHe09}. We present results
both for one-point estimators and also generalise them to two corresponding power spectra.
\textsection6 is devoted to adding relevant linear correction terms to these estimators
in the absence of spherical symmetry. Next, in \textsection7 we introduce inverse covariance weighting and
design a trispectrum estimator which is optimal for arbitrary scanning strategy. The estimator
used in this section will also be useful also in estimating secondary non-Gaussianity. For homogeneous 
noise and all-sky coverage the estimators are identical. Finally, \textsection8
is devoted to concluding discussions and future prospects. 

\section{Correlation functions and the Cumulant Correlators}
\label{sec:intro}
The temperature fluctuations of the Cosmic Microwave Background (CMB)
are typically assumed to be a realisation of statistically isotropic 
Gaussian random field. For Gaussian perturbations all the information
needed to provide a complete statistical description is contained in the
power spectrum of the distribution; for a non-Gaussian
distribution higher-order correlation functions are also needed. 
With the assumption of {\em weak} non-Gaussianity
only the first few correlation functions are needed to describe the
departure from Gaussianity. We will denote the n-point correlation function by
$\xi_N(\oh_1,\dots,\oh_n)$. The n-point correlation functions are decomposed
into parts which are purely Gaussian in nature and those which signify
departures from Gaussianity. These are also known as {\em connected} and {\em disconnected}
terms because of their representation by respective diagrams; see \cite{BernardReview} for more details.
At the level of the four-point correlation  function the corresponding connected part,
denoted by the subscript $\langle \dots \rangle_c$), $\mu_4$ can be defined as:

\begin{equation}
\mu_4(\oh_1,\dots,\oh_4) = 
\langle \delta T(\oh_1) \dots \delta T(\oh_4)\rangle_c.
\end{equation}

\n
The connected component of the four-point function will be exactly zero
for a purely Gaussian temperature field. The Gaussian contribution on the
other hand can be written as a product of two two-point correlation
functions. So the total four-point correlation function can be
written as the sum of the connected and the disconnected part:

\begin{equation}
\xi_4(\oh_1\dots\oh_4) = \xi_2(\oh_1,\oh_2)\xi_2(\oh_3,\oh_4) + 
\xi_2(\oh_1,\oh_3)\xi_2(\oh_2,\oh_4) + 
\xi_2(\oh_1,\oh_4))\xi_2(\oh_2,\oh_3) + \mu_4(\oh_1\dots\oh_4).
\end{equation}

\n
As we will see, the Gaussian part will add to the scatter associated with
any estimator for the four-point correlation function. At lower level
$\xi_2(\oh_1,\oh_2)=\mu_2(\oh_1,\oh_2)$ and $\xi_3(\oh_1\dots\oh_3)=\mu_3(\oh_1\dots\oh_3)$ and hence there are no disconnected parts. The number of 
degrees of freedom associated with higher-order correlations increase
exponentially with its order. This is mainly due to the increased number 
of configurations possible for which one can measure a higher-order
correlation function. The cumulant correlators are defined by identifying
all available vertices or points to just two points. There are
two such cumulant correlators which can constructed at the four-point level.

\begin{equation}
\xi_{31} \equiv \langle \delta^3 T(\oh_1)\delta T(\oh_2) \rangle_c; \qtwo
\xi_{22} \equiv \langle \delta^2 T(\oh_1)\delta T^2 (\oh_2) \rangle_c.
\label{eq:cumu}
\end{equation}

\n
These two degenerate sets of cumulant correlators carry 
information at the level of four-point, but they are essentially two-point
correlation functions and can be studied in the harmonic domain by
their associated power spectra. The first, $\xi_{31}$
was studied in the context of 21cm surveys by \cite{Coo21} and \cite{Cooray8} and
the second $\xi_{22}$, was shown to have the power to separate 
the lensing contribution from the Kinetic Sunya\'ev Zeldovich
effect \citep{RiqSper}. These are natural generalisations of their third-order
counterpart recently studied by \cite{MuHe09}, who introduced their
 optimised form for direct use on realistic data. These were later used by 
\cite{Smidt09} to estimate $f_{NL}$  and \cite{Calab09} to study lensing-secondary correlations from
WMAP5 data.

Although there are various advantages of working in real space, it is often
easier to work in the harmonic domain. The main motivation to work in the harmonic domain
is linked to the fact that inflationary models predict a well-defined
peak structure for the power spectrum. These structures
are well-known diagnostics for constraining cosmology at various levels. This is
true for higher-order multi-spectra as they also involve the effect of the transfer functions.
Note that the noise in CMB experiments is typically assumed to be Gaussian and will therefore
contribute only to the disconnected terms.

\section{Fourier Transforms of Cumulant Correlators and their Optimum Estimators}

The real-space correlation functions are clearly very important tools which can
be used in surveys with patchy sky coverage. However recent CMB surveys scan
the sky with near all-sky coverage. This makes a harmonic-space description more
appropriate as various symmetries can be included in a more straightforward way.
For Gaussian random fields with all-sky coverage the estimates of various statistics are loosely speaking uncorrelated. 
Even for a non-Gaussian field, they are reasonably uncorrelated at different angular scales or at different $l$ as 
long as the non-Gaussianity is weak. We begin by introducing the harmonic transform of the observed temperature map 
$\delta T(\oh)$ ($\hat \Omega = (\theta,\phi)$) for all-sky coverage:

\begin{equation}
a_{lm} \equiv \int d\hat\Omega \frac{\Delta T(\hat\Omega)}{T} Y_{lm}^*(\hat\Omega)
\equiv \int d\hat\Omega {\delta T(\hat\Omega)}Y_{lm}^*(\hat\Omega).
\end{equation}

\n
Realistically however, we will only be observing the part of the sky which is not
masked by Galactic foregrounds. The window function $w(\Omega)$ which we will take
as a completely general window can be used to define what is known as the pseudo harmonics
which we designate as $\tilde a_{lm}$.

\begin{equation}
\tilde a_{lm} \equiv 
\int d\hat\Omega w(\hat\Omega) {\delta T(\hat\Omega)}Y_{lm}^*(\hat\Omega).
\end{equation}

\n
We follow this description for rest of this paper. Any statistic $X$ obtained from
the masked sky will be denoted as $\tilde X$ and the estimated all-sky version will be
denoted $\hat X$. The ensemble averages of the unbiased all-sky estimators which coincide
with the theoretical models will be denoted only by the corresponding latin symbol $X$.

\subsection{Two-Point Estimators for the Trispectrum $C_l^{(3,1)}$, $C_l^{(2,2)}$}

\noindent
At the level of four point cumulant correlators we have two different estimators 
which can independently be used to study the trispectrum. These estimators that we discuss
here are direct harmonics transforms of these two point correlators. We introduced 
the cumulant correlators in Eq.(\ref{eq:cumu}). The two estimators we define in this section are related to
$\langle \delta T^2(\oh)\delta^2 T(\oh')\rangle$ and $\langle \delta T^3(\oh)\delta T(\oh')\rangle$ 
through harmonic transforms.
For the cubic function $\delta T^3(\oh)$, we denote the harmonic transform as $a^{(3)}_{lm}$
and similarly $a^{(2)}_{lm}$ is the harmonic transform of the quadratic form $\delta T^2(\oh)$.
We obtain the following relations for all-sky coverage,
and for the pseudo-harmonics $\tilde a^{(2)}_{lm}$ with a mask.

\ben
&& a^{(2)}_{lm}  = \sum_{l_1m_1}\sum_{l_2m_2} a_{l_1m_1}a_{l_2m_2}
\int d \hat \Omega Y_{l_1m_1}(\hat \Omega)Y_{l_2m_2}(\hat \Omega) Y^*_{lm}(\hat \Omega);  \\
&& \tilde a^{(2)}_{lm}  =  \sum_{l_1m_1} \dots \sum_{l_3m_3} a_{l_1m_1}a_{l_2m_2}w_{l_3m_3} 
\int d \hat \Omega Y_{l_1m_1}(\hat \Omega)Y_{l_2m_2}(\hat \Omega)Y_{l_3m_3}(\hat \Omega)Y^*_{\
lm}(\hat \Omega)  = \sum_{(l'm')}K_{lml'm'} a^{(2)}_{l'm'}. 
\een

\n
The corresponding results for $a^{(3)}_{lm}$ and $\tilde a^{(3)}_{lm}$ are as follows:

\ben
&& a^{(3)}_{lm} = \sum_{l_1m_1} \dots \sum_{l_3m_3}  a_{l_1m_1}a_{l_2m_2}a_{l_3m_3}
\int d \hat \Omega Y_{l_1m_1}(\hat \Omega)Y_{l_2m_2}(\hat \Omega)Y_{l_3m_3}(\hat \Omega) Y^*_{\
lm}(\hat \Omega); \\
&& \tilde a^{(3)}_{lm} = \sum_{l_1m_1} \dots \sum_{l_4m_4} a_{l_1m_1}a_{l_2m_2}a_{l_3m\
_3}w_{l_4m_4}
\int d \hat \Omega Y_{l_1m_1}(\hat \Omega)Y_{l_2m_2}(\hat \Omega)Y_{l_3m_3}(\hat \Omega)Y_{l_4m_4}(\hat \Omega) Y^*_{\
lm}(\hat \Omega)   = \sum_{(l'm')}K_{lml'm'} a^{(3)}_{l'm'}.
\een

\n
The coupling matrix $K_{lml'm'}$ encodes information about the mode coupling which is 
introduced because of the masking of the sky \citep{Hiv}:

\begin{equation}
K_{l_1m_1l_2m_2}[w] = \int w({\on}) Y_{l_1m_1}(\on) Y_{l_2m_2}(\on) {\rm d} \on 
=\sum_{l_3m_3} \tilde w_{l_3m_3} \left( {(2l_1+1)(2l_2+1)(2l_3+1)\over 4\pi } \right)^{1/2} 
\left ( \begin{array}{ c c c }
     l_1 & l_2 & l_3 \\
     0 & 0 & 0
  \end{array} \right)
\left ( \begin{array}{ c c c }
     l_1 & l_2 & l_3 \\
     m_1 & m_2 & m_3
  \end{array} \right).
\end{equation}

\n
The matrices here denote the 3J symbols \citep{Ed68}. Using the harmonic transforms we define the following 
power spectra which can directly probe the trispectra. 

\begin{equation}
\myC_l^{(2,2)} = {1 \over 2l+1} \sum_m a_{lm}^{(2)*} a_{lm}^{(2)};  ~~~~~
\tilde \myC_l^{(2,2)} = {1 \over 2l+1} \sum_m  \tilde a_{lm}^{(2)*} \tilde a_{lm}^{(2)}.
\end{equation}

\n
From the consideration of isotropy and homogeneity we can write the following relations:

\begin{equation}
\langle a_{lm}^{(2)} a_{lm}^{(2)*}\rangle_c = \myC_l^{(2,2)} \delta_{ll'}\delta_{mm'}; \qtwo
\langle a_{lm}^{(3)} a_{lm}^{*}\rangle_c = \myC_l^{(3,1)} \delta_{ll'}\delta_{mm'}.
\end{equation}

\n
The pseudo-power spectrum which is recovered from the masked harmonics is defined
in an analogous way. Finally the resulting power spectra $C_l^{(2,2)}$ can be expressed 
in terms of the trispectrum $T_{l_3l_4}^{l_1l_2}(l)$ by the following expression:

\begin{equation}
\myC_l^{(2,2)} = \sum_{l_1l_2l_3l_4} T^{l_3l_4}_{l_1l_2}(l) \sqrt {(2l_1+1)(2l_2+1)\over 4\pi (2l+1)}
\sqrt {(2l_3+1)(2l_4+1)\over 4\pi (2l+1)}
\left ( \begin{array}{ c c c }
     l_1 & l_2 & l \\
     0 & 0 & 0
  \end{array} \right)
\left ( \begin{array}{ c c c }
     l_3 & l_4 & l \\
     0 & 0 & 0
  \end{array} \right).
\end{equation}

\n
The trispectrum which is introduced here is a four-point correlation function in harmonic space.
The definition here ensures that it is invariant under various transformations; see \cite{Hu00} and \cite{huOka02} for detailed discussion:

\begin{equation}
\langle a_{l_1m_1} a_{l_2m_2} a_{l_3m_3} a_{l_4m_4} \rangle_c =
\sum_{LM} (-1)^M T_{l_1l_2}^{l_3l_4}(L) \left ( \begin{array}{ c c c }
     l_1 & l_2 & L \\
     m_1 & m_2 & M
  \end{array} \right)
\left ( \begin{array}{ c c c }
     l_3 & l_4 & L \\
     m_3 & m_4 & -M
  \end{array} \right).
\end{equation}

\noindent
Partial sky coverage can be dealt with in an exactly similar manner. By using the
harmonic transforms of the masked sky and expressing the masked harmonics in terms of 
all-sky harmonics we can relate the PCL versions of these power spectra $\tilde C_l^{(2,2)}$
in terms of its all-sky components  $C_l^{(2,2)}$. This involves the
harmonic transform of the mask $W_{lm}$ and its associated power spectrum  $W_{l}$.
The coupling matrix for the  $M_{ll'}$ is the same as that which is used for 
inverting the PCL power spectrum to recover the unbiased power spectrum.

\ben
{\tilde {\cal C}}_l^{(2,2)} &=&  \sum_{l'l''} (2l'+1)
\left ( \begin{array}{ c c c }
     l & l' & l'' \\
     0 & 0 & 0
  \end{array} \right)^2
{ (2l'' +1 )\over 4\pi} |w_{l''}^2| \nonumber \\
&&\times\sum_{l_1l_2l_3L} T^{l_1l_2}_{l_3l_4}(l') \sqrt {(2l_1+1)(2l_2+1)\over 4\pi (2l'+1)\
}
\sqrt {(2l_3+1)(2l_4+1)\over 4\pi (2l'+1)}
\left ( \begin{array}{ c c c }
     l_1 & l_2 & l' \\
     0 & 0 & 0
  \end{array} \right)
\left ( \begin{array}{ c c c }
     l_3 & l_4 & l' \\
     0 & 0 & 0
  \end{array} \right) = \sum_{l'} M_{ll'}{\cal C}_{l'}^{(2,2)}.
\een

\n
where
\begin{equation}
M_{ll'} = (2l'+1)\sum_{l''}
\left ( \begin{array}{ c c c }
     l & l' & l'' \\
     0 & 0 & 0
  \end{array} \right)^2
{ (2l'' +1 )\over 4\pi} |w_{l''}^2|.
\end{equation}
\noindent
In the literature the power spectrum $c_l^{(2,2)}$ is also referred as the {\em second spectrum}.
The other degenerate cumulant correlator of the same order that contains information about
the trispectrum can be written as:

\begin{equation}
\myC_l^{(3,1)} = {1 \over 2l+1} \sum_m \mathrm {Real} \left \{ a_{lm}^{(3)*} a_{lm}^{(1)} \right \} ;~~~~~
\tilde \myC_l^{(3,1)} = {1 \over 2l+1} \sum_m \mathrm {Real} \left \{  \tilde a_{lm}^{(3)*} \tilde a_{lm}^{(1)} \right \}.
\end{equation}

\n
The resulting power spectrum which probes the various components of the trispectrum with different weights
can be expressed as follows:

\begin{equation}
{\cal C}_l^{(3,1)} = \sum_{l_1l_2l_3L} T^{l_1l_2}_{l_3l}(L) \sqrt {(2l_1+1)(2l_2+1)\over 4\pi \
(2L+1)}
\sqrt {(2L+1)(2l_3+1)\over 4\pi (2l+1)}
\left ( \begin{array}{ c c c }
     l_1 & l_2 & L \\
     0 & 0 & 0
  \end{array} \right)
\left ( \begin{array}{ c c c }
     L & l_3 & l \\
     0 & 0 & 0
  \end{array} \right).
\end{equation}

\noindent
With partial sky coverage we can proceed as before to connect the PCL version of the estimator ${\cal C}_l^{(3,1)}$
to its all-sky analogue. The resulting estimators combine the values of various components of the trispectrum 
with different weighting. The associated power-spectrum will therefore have a different dependence on
parameters describing the primordial trispectrum.

\begin{eqnarray}
{\cal C}_l^{(3,1)} &=&  \sum_{l'l''} (2l'+1)
\left ( \begin{array}{ c c c }
     l & l' & l'' \\
     0 & 0 & 0
  \end{array} \right)^2
{ (2l'' +1 )\over 4\pi} |w_{l''}^2| \nonumber \\
&&\times\sum_{l_1l_2l_3L} T^{l_1l_2}_{l_3l}(L) \sqrt {(2l_1+1)(2l_2+1)\over 4\pi (2L+1)\
}
\sqrt {(2L+1)(2l_3+1)\over 4\pi (2l'+1)}
\left ( \begin{array}{ c c c }
     l_1 & l_2 & L \\
     0 & 0 & 0
  \end{array} \right)
\left ( \begin{array}{ c c c }
     L & l_3 & l' \\
     0 & 0 & 0
  \end{array} \right) = \sum_{l'} M_{ll'}{\cal C}_{l'}^{(3,1)}.
\end{eqnarray}

\n
The  links to real space cumulant correlators are same as their third order counterpart:

\be
{\langle \delta^2 T(\oh) \delta^2 T(\oh')\rangle_c} = {1 \over 4 \pi} \sum_l (2l+1)P_l(\cos(\oh\cdot\oh')){\cal C}_l^{(2,2)} ; \qtwo
{\langle \delta^3 T(\oh) \delta T(\oh')\rangle_c} = {1 \over 4 \pi} \sum_l (2l+1)P_l(\cos(\oh\cdot\oh')){\cal C}_l^{(3,1)} .
\ee

\n
Hence we can conclude that it is possible to generalise these results
to an arbitrary mask with arbitrary weighting functions.
The deconvolved set of estimators at order $p,q$ can be written as follows:

\be
{\hat C}_l^{p,q} = M_{ll'}^{-1} {\tilde C}_{l'}^{p,q}.
\ee

\n
{\em This is one of the important results of this paper}.
The mask used for various orders $p,q$ is the same, but with the increasing order
the number of degenerate power-spectra that can be constructed from a multi-spectrum
increases drastically. As we have seen at the level of bispectrum we can keep one
of the triangle sides fixed and sum over all contributions from all possible configurations of
the triangle. Similarly for the trispectrum we can keep one of the sides of the 
rectangle fixed, or one of the diagonals fixed, and sum over all possible configurations.
The possibilities increase as we move to higher-order in multispectra. Another related
complication would be from the number of disconnected terms which need to be subtracted as 
they simply correspond to Gaussian contributions. At the 4-point level we 
need to subtract the disconnected pieces from the dominant Gaussian component of temperature fluctuations
including the noise \citep{Hu00,huOka02}:

\be
G^{l_1l_2}_{l_3l_4}(L) =  (-1)^{l_1+l_3} \sqrt {(2l_1+1)(2l_3+1)} C_{l_1}C_{l_3} \delta_{L0}\delta_{l_1l_2}\delta_{l_2l_3}
  + (2L+1)C_{l_1}C_{l_2}\big [ (-1)^{l_2+l_3+L} \delta_{l_1l_3}\delta_{l_2l_4} + \delta_{l_1l_4}\delta_{l_2l_3}\big ],
\ee
where $C_l$ is the power spectrum including noise.

For the all-sky case and if we restrict only to modes with ordering $l_1 \le l_2 \le l_3 \le l_4$, 
the non-zero component corresponds to terms with $L=0$ or $l_1=l_2=l_3=l_4$.
However for arbitrary sky coverage, which results in mode-mode coupling, no such
general comments can be made. Estimators developed here in their all-sky form are known in the literature, and 
the results here show how to generalise them for partial sky coverage. However
the main interest in data compression lies in using optimal weights for the
compression, which is model-dependent as
the weights depend on the specific model being probed.

If we introduce the Gaussian contributions to $C_l^{(3,1)}$ as $G_l^{(3,1)}$, and to
$C_l^{(2,2)}$ as $G_l^{(2,2)}$, in the presence of partial sky coverage we can write:

\be
\tilde G_l^{(3,1)} = \sum_{l'} M_{ll'}G_{l'}^{(3,1)} ; \qtwo \tilde G_l^{(2,2)} = \sum_{l'}  M_{ll'}G_{l'}^{(2,2)}.
\ee

\n
For  all-sky coverage one can obtain the corresponding results by replacing $T^{l_1,l_2}_{l_3,l_4}(L)$
by $G^{l_1,l_2}_{l_3,l_4}(L)$ in respective equations:

\begin{equation}
{ G}_l^{(3,1)} = \sum_{l_1l_2l_3L} G^{l_1,l_2}_{l_3,l}(L) \sqrt {(2l_1+1)(2l_2+1)\over 4\pi \
(2L+1)}
\sqrt {(2L+1)(2l_3+1)\over 4\pi (2l+1)}
\left ( \begin{array}{ c c c }
     l_1 & l_2 & L \\
     0 & 0 & 0
  \end{array} \right)
\left ( \begin{array}{ c c c }
     L & l_3 & l \\
     0 & 0 & 0
  \end{array} \right)
\end{equation}

\begin{equation}
G_l^{(2,2)} = \sum_{l_1l_2l_3L} G^{l_3,l_4}_{l_1,l_2}(l) \sqrt {(2l_1+1)(2l_2+1)\over 4\pi (2l+1)}
\sqrt {(2l_3+1)(2l_4+1)\over 4\pi (2l+1)}
\left ( \begin{array}{ c c c }
     l_1 & l_2 & l \\
     0 & 0 & 0
  \end{array} \right)
\left ( \begin{array}{ c c c }
     l_3 & l_4 & l \\
     0 & 0 & 0
  \end{array} \right).
\end{equation}

To subtract the disconnected Gaussian part, we use simulations. These are constructed from
the same power spectrum that describes the non-Gaussian maps. Noise realisations which describe
the actual map are also introduced in the Gaussian maps and exactly the same mask is used.
This will ensure that the estimator remains unbiased. If we denote the final estimators after 
the subtraction of Gaussian disconnected parts by $D_l^{3,1}$ and $D_l^{(2,2)}$ we can write for 
all-sky coverage:

\be
\hat D_l^{(2,2)} \equiv \hat C_l^{(2,2)} - \hat G_l^{(2,2)}; \qtwo \hat D_l^{(3,1)} \equiv \hat C_l^{(3,1)} - \hat G_l^{(3,1)}.
\ee

\n
In the presence of a mask, deconvolved estimators are related to the masked estimators
by the mixing matrix $M_{ll'}$:

\be
\hat D_l^{(2,2)} = \sum_{ll'} M^{-1}_{ll'}(\tilde C_{l'}^{(2,2)} - \tilde G_{l'}^{(2,2)}); \qtwo 
\hat D_l^{(3,1)} = \sum_{ll'} M^{-1}_{ll'}(\tilde C_{l'}^{(3,1)} - \tilde G_{l'}^{(3,1)}).
\ee

Practical implementation of these estimators can provide a quick sanity check of a pipeline
design for non-Gaussian estimators. While it is useful to keep in mind that these
estimators are sub-optimal they are nevertheless {\em unbiased} and, depending on various 
choices of $f_{NL}$ and $g_{NL}$, they can provide an analytical basis for
computation of the scatter and cross-correlation among various estimators associated with
different levels of the correlation hierarchy. 

The dependence of  $D_l^{(2,2)}$ and  $D_l^{(3,1)}$ on $f_{NL}$ and $g_{NL}$ are
different and hence can be used to provide independent constraints on both of
these parameters without using third-order estimators and might be useful for
providing cross-checks.

\subsection{One-Point Unoptimised Estimators for the Trispectrum $C_l^{(4)}$}

\n
The one-point cumulants at third order can be written in terms of the bispectra as:

\begin{equation}
\mu_3 = \langle \delta T^3 (\ho)\rangle = {1\over 4\pi}\int \delta T^3 (\ho) d\ho = {1 \over 4 \pi} \sum_{l_1l_2l_3} h_{l_1l_2l_3} B_{l_1l_2l_3}.
\end{equation}

\n 
Similarly the one-point cumulant at fourth order can be written in terms as the trispectra as:

\begin{equation}
\mu_4 = \langle \delta T^4 (\ho)\rangle_c  = {1 \over 4\pi}\int \delta T^4 (\ho) d\ho = { 1\over 4\pi} \sum_L \sum_{l_1l_2l_3l_4} h_{l_1l_2L}h_{l_3l_4L}
 T_{l_1l_2}^{l_3l_4}(L).
\end{equation}

\n
We can also define the $S_N$ parameters generally used in the literature as: 

\be
S_3 = \mu_3; \qtwo S_4 = \mu_4 - 3\mu_2^2.
\ee

\n
Throughout this discussion we have absorbed the experimental beam  $b_l = \exp\{ - l(l+1)/2\sigma_b^2 \}$ 
in the harmonics $a_{lm}$ unless it is displayed explicitly. Here $\sigma_b = {\rm FWHM}/\sqrt {8ln2}$ for a gaussian beam. Alternatively it is also possible to define by respective powers of $\mu_2$ to make these one-point estimators
less sensitive to the normalisation of the power spectra. In that case we will have $S_3 = {\mu_3 / \mu_2^2}$.
These moments at fourth order are generalisations of the third-order moments, used as a basis for 
the construction of  estimators for $f_{\rm NL}$ by introducing the optimal weights  $A(r,\oh)$ and $B(r,\oh)$.

\section{Models for Primordial Non-Gaussianity and Construction of Optimal Weights}

The optimisation techniques that we introduce in next section follow the discussion in \cite{MuHe09}.
The optimisation procedure depends on construction of three-dimensional fields from the
harmonic components of the temperature fields $a_{lm}$ with suitable weighting with respective
functions which describes primordial non-Gaussianity\citep{YKW}. These weights make the estimators
act optimum and the matched filtering technique adopted ensures that the response to the observed non-Gaussianity is 
maximum when it matches with primordial non-gaussianity corresponding to the weights.

In the linear regime the curvature perturbations which generate the fluctuations in the CMB sky are
written as:

\begin{equation}
a_{lm} = 4\pi (-i)^l \int {d^3k \over (2\pi)^3} \Phi({\bf k}) \Delta_l^T(k)Y_{lm}(\hat k).
\end{equation}
\n

\n
We will need the following functions to construct the harmonic space trispectra as well
as to generate weights for construction of optimal estimators (for a more complete description
of predicting trispectrum from a given Inflationary prediction see \citep{Hu00,huOka02}) :

\be
\alpha_l(r) = {2 \over \pi} \int_0^{\infty} k^2 dk \Delta_l(k)j_l(kr); ~~
\beta_l(r) = {2 \over \pi} \int_0^{\infty} k^2 dk P_{\phi}(k) \Delta_l(k)j_l(kr); ~~
\mu_l(r) = {4 \over \pi} \int_0^{\infty} k^2 dk \Delta_l(k)j_l(kr) 
\ee

In the limit of low multipoles where the perturbations are mainly dominated by Sachs-Wolfe Effect
the transfer functions  $\Delta_l(k)$ takes a rather simple form $\Delta_l(k)= 1/3 j_l(kr_*)$
where $r_* = (\eta_0 - \eta_{dec})$ denotes the time elapsed between cosmic recombination
and the present epoch. In general the transfer function needs to be computed numerically.
The {\em local} Model for the primordial bispectrum and trispectrum can be constructed by going beyond linear theory
in the expansion of the $\Phi(x)$. Additional parameters $f_{NL}$ and $g_{NL}$ are introduced 
which need to be estimated from observation. As discussed in the introduction, $g_{NL}$ can be linked to $r$ the
scalar to tensor ratio in a specific inflationary model and hence expected to be small.

\begin{equation}
\Phi({\bf x}) = \Phi_L({\bf x}) + f_{NL} \left ( \Phi^2_L({\bf x}) - \langle \Phi_L^2({\bf x}) \rangle \right ) + g_{NL} \Phi_L^3({\bf x})
+ h_{NL} \left ( \Phi^4_L({\bf x}) - 3\langle \Phi_L^2({\bf x}) \rangle \right ) + \dots
\end{equation}

We will only consider the local model of primordial non-Gaussianity in this paper and only adiabatic
initial perturbations. More complicated cases of primordial non-Gaussianity will be dealt with elsewhere.
In terms of inflationary potential $V(\phi)$ associated with a scalar potential $\phi$ one can express
these constants as \citep{Hu00}:

\be
f_{\rm NL} = -{5 \over 6} {1 \over 8\pi G} {\partial^2 \ln V(\phi) \over \partial \phi^2}; \qtwo
g_{\rm NL} = {25 \over  54} {1 \over  8 \pi G} \left [ 2\left ( {\partial^2 \ln V(\phi) \over \partial \phi^2}\right )
- \left ( \partial^3 \ln V(\phi) \over \partial \phi^3 \right ) \left ( \partial \ln V(\phi) \over \partial \phi\right ) \right ].
\ee

There are two contribution to primordial non-Gaussianity. The first part is parametrised by 
$f_{\rm NL}$ and the second contribution 
is proportional to a new parameter which appears at fourth order which we denote by $g_{\rm NL}$.
From theoretical considerations in generic models of inflation one would expect $g_{\rm NL} \le r /50 $
with $r$ being the scalar to tensor ratio \citep{Seery07}.
 
\n
Following \citep{Hu00} we can expand above expression in Fourier space to write:

\begin{eqnarray}
&&\Phi_2(k) = \int {d^3\bk_1 \over (2\pi)^3} \Phi_L(\bk+\bk_1)\Phi_L^*(\bk_1) - (2\pi)^3 \delta_D(\bk)\int {d^3\bk_1 \over (2\pi)^3} P_{\Phi\Phi}(\bk_1) \\
&&\Phi_3(k) = \int {d^3\bk \over (2\pi)^3} \Phi_L(\bk_1)\Phi_l(\bk_2)\Phi_l^*(\bk_1).
\end{eqnarray}

\n
The resulting trispectra associated with these perturbations can be expressed as:

\begin{equation}
T_{\Phi}(k_1,k_2,k_3,k_4) \equiv \langle \delta(\bk_1)\delta(\bk_2)\delta(\bk_3)\delta(\bk_4)\rangle_c
= \int {d^3{\bK} \over (2 \pi)^3} \delta_D(\bk_1 +\bk_2 -\bK) \delta_D(\bk_3+\bk_4+\bK) 
{\cal T}_{\Phi}(\bk_1,\bk_2,\bk_3,\bk_4,\bK).
\end{equation}

\n
where the ${\cal T}(\bk_1,\bk_2,\bk_3,\bk_4,\bK)$ can be decomposed into two different constituents.

\begin{eqnarray}
&& {\cal T}_{\Phi}^{(2)}(\bk_1,\bk_2,\bk_3,\bk_4,\bK) = 4 f_{NL}^2 P_{\phi}(\bK) P_{\phi}(\bk_1) P_{\Phi}(\bk_3) \\
&& {\cal T}_{\Phi}^{(3)}(\bk_1,\bk_2,\bk_3,\bk_4,\bK) = g_{NL} \left \{ P_{\phi}(\bK) P_{\phi}(\bk_1) P_{\Phi}(\bk_3) +
P_{\phi}(\bK) P_{\phi}(\bk_1) P_{\Phi}(\bk_3) 
\right \}
\end{eqnarray}

\n
The CMB trispectrum now can be written as

\ben
T^{l_1l_2}_{l_3l_4}(L) =&& 4f_{\rm NL}^2 h_{l_1l_2L}h_{l_3l_4L}\int r_1^2dr_1 r_2^2 dr_2 F_L(r_1,r_2) \alpha_{l_1}(r_1)\beta_{l_2}(r_1)\alpha_{l_3}(r_2)\beta_{l_4}(r_2)
\nonumber \\
&& + g_{\rm NL} h_{l_1l_2L}h_{l_3l_4L}\int r^2 dr \beta_{l_2}(r) \beta_{l_4}(r) [ \mu_{l_1}(r)\beta_{l_3}(r) + \mu_{l_3}(r)\beta_{l_1}(r)].
\een

For detailed descriptions involving polarisation maps see \citep{huOka02,Hu00,KomSpe01,Kogo06}.
The CMB bispectrum which describes departures from Gaussianity at the lowest level can be similarly written as:

\be
B_{l_1l_2l_3} = 2f_{\rm NL} h_{l_1l_2l_3}\int r^2 dr \left [ \alpha_{l_1}(r) \beta_{l_2}(r)\beta_{l_3}(r) + \alpha_{l_2}(r) \beta_{l_3}(r)\beta_{l_1}(r) + \alpha_{l_3}(r) \beta_{l_2}(r)\beta_{l_1}(r)  \right ].
\ee

\n
We have defined the following form factor to simplify the display:

\be
h_{l_1l_2l_3} = \sqrt{(2l_1+1)(2l_2+1)(2l_3+1)\over 4 \pi}
\left ( \begin{array}{ c c c }
     l_1 & l_2 & l_3 \\
     0 & 0 & 0
  \end{array} \right). \\
\ee

The extension to order beyond presented here to involve higher-order Taylor coefficients may not be practically useful as the detector noise and 
the cosmic variance may prohibit any reasonable signal-to-noise.

\section{Near Optimal Estimators for Trispectrum}

\n
We will develop two estimators for extraction of power spectra associated with the trispectrum \citep{Hu00,huOka02} in this section
$\myK_l^{3,1)}$ and $\myK_l^{2,2}$. These estimators are optimised version of similar estimators considered
in the previous section. The derivation here parallels that of the previous section where unoptimised versions of 
these estimators were developed which uses the PCL type estimators. We will start by introducing four 
different fields which are constructed from the temperature fields. These fields 
are defined over the observed part of the sky and and are constructed using suitable
weights to temperature harmonics. These weights are functions of radial distance $\alpha_l(r),\beta_l(r), \mu_l(r)$ 
so the constructed fields are 3D fields. This method follows the same technique as introduced by
\cite{KSW}.

\ben
&& A(r,\oh) = \sum_{lm} A_{lm}Y_{lm}(\oh); \qquad B(r,\oh) = \sum_{lm} B_{lm}Y_{lm}(\oh); \qquad  M(r,\oh) = \sum_{lm} M_{lm}Y_{lm}(\oh); \\
&& A_{lm} =  {\alpha_{lm} \over C_l} a_{lm}; ~~ B_{lm} =  {\beta_{lm} \over C_l} a_{lm}; ~~ M_{lm} =  {\mu_{lm} \over C_l} a_{lm} .
\een

\n
We have absorbed the beam smoothing into the corresponding harmonic coefficients. In addition to the weighting
functions we will also need to define overlap integrals $F_L(r_1,r_2)$ which act as a kernel 
for cross-correlating fields at two different radial distances. Note that
the overlap integral depends on the quantum number $L$:

\be
F_L(r_1,r_2) = {2 \over \pi} \int k^2 dk P_{\Phi\Phi}(k) j_L(kr_1)j_L(kr_2).
\ee

\n
In the following subsections we will use specific forms for the trispectra to construct optimal 
estimator. Although the estimators developed here are specific to a given model clearly for any given
model for the trispectra we can obtain similar construction in an optimal way. The weighting of these
harmonics make the estimator a match-filtering one which ensures maximum response when the trispectra
obtained from the data matches with that with theoretical construct. Inverse variance weighting
ensures that the estimator remains optimal for a specific survey strategy.

\subsection{Estimator for $\myK^{(4)}$}

\n
The one-point estimators are simpler though they compress all available information in one number.
One-point estimators can also be computed directly in real space and hence are simpler to compute 
when dealing with partial sky coverage in the presence of inhomogeneous noise.
Using notations introduced above, we can write the one-point cumulant at fourth order by the following expression:

\be
\myK^{(4)} = 4f_{\rm Nl}^2 \int r_1^2 \int r_2^2 dr_2 F(r_1,r_2) \int d\oh A(r_1,\oh)B(r_1,\oh)B(r_2,\oh)A(r_2,\oh) 
         + 2g_{\rm Nl} \int r^2 dr \int d\oh A(r,\oh)B^2(r,\oh)M(r,\oh)
\ee

\n
which can be written in a compact form

\be
\myK^{(4)} = 4f_{\rm NL}^2  \int r_1^2 \int r_2^2 dr_2 F(r_1,r_2) \langle A(r_1,\oh)B(r_1,\oh)A(r_2,\oh)B(r_2,\oh) \rangle
 + 2g_{\rm Nl} \int r^2 dr \langle A(r,\oh)B^2(r,\oh)M(r,\oh) \rangle.
\ee

\n
As expected there are two parts in the contribution. The first part depends on two radial directions
through the overlap integral $F(r_1,r_2)$;  the second is much simpler and just contains one
line-of-sight integration. The amplitude of the first term depends on $f_{\rm NL}^2$ and the 
second term  is proportional to $g_{\rm NL}$. Typically they contribute 
equally to the resulting estimate.

\subsection{Estimator for $\myK_l^{(3,1)}$}

\n
Moving beyond the one-point cumulants we can construct the estimators of the two power spectra which
we discussed before, $C_l^{(3,1)}$ and $C_l^{(2,2)}$.  The corresponding optimised versions
can be constructed by cross-correlating the fields $A(r_1,\oh)B(r_1,\oh)B(r_2,\oh)$ with
$B(r_2,\oh)$. Clearly in the first case the harmonics depend on two radial distances $r_1,r_2$ for 
any given angular direction:

\be
A(r_1)B(r_1)B(r_2)|_{lm} = \int A(r_1,\oh)B(r_1,\oh)B(r_2,\oh)~Y_{lm}(\oh)~d\oh; \qquad
A(r_2)|_{lm} = \int A(r_2,\oh)~Y_{lm}(\oh)~d\oh.
\ee

\n
Next we compute the  cross-power spectra ${\cal J}_l^{AB^2,A}(r_1,r_2)$ which also depend on both radial
distances $r_1$ and $r_2$:

\be
 {\cal J}_l^{AB^2,A}(r_1,r_2) = \frac{1}{2l+1} \sum_m F_{L}(r_1,r_2) {\re} \left [ \{A(r_1)B(r_1)B(r_2) \}_{lm} \{ A(r_2) 
\}_{lm}^* \right ].  \\
\ee

\n
The construction for the second term is very similar. We start by decomposing the real space product $A(r,\oh)B^2(r,\oh)$
and $M(r,\oh)$ in harmonic space. There is only one radial distance involved in both of these terms.

\be
A(r)B^2(r)|_{lm} = \int [A(r,\oh)B^2(r,\oh)]~Y_{lm}(\oh)~d\oh; \qquad
M(r,\oh)|_{lm} = \int M(r) Y_{lm}(\oh)~d\oh.
\ee

\n
Finally the line-of-sight integral which involves two overlapping contribution through the weighting
kernels for the first term and only one for the second gives us the required estimator:

\be
{\cal K}_l^{(3,1)} = 4f_{\rm nl}^2 \int r_1^2 dr_1 \int r_2^2 dr_2  {\cal J}_l^{AB^2,A}(r_1,r_2) + 2g_{\rm nl} \int r^2 dr {\cal L}_l^{AB^2,M}(r)
\label{eq:k31}
\ee

Next we show that the construction described above does reduces to an optimum estimator for the power spectrum
associated with the trispectrum. The harmonics associated with the product field $A(r_1)B(r_1)B(r_2)$ can be 
expressed in terms of the functions $\alpha(r)$ and $\beta(r)$:

\be
A(r_1)B(r_1)B(r_2)|_{lm} = \sum_{LM} (-1)^M \sum_{lm,l_im_i} a_{l_1m_1}a_{l_2m_2}a_{l_3m_3} ~
\alpha_{l_1}(r_1)\beta_{l_2}(r_1)\beta_{l_3}(r_2) {\cal G}_{l_1l_2L}^{m_1m_2M}{\cal G}_{Ll_3l}^{Mm_3m}.  \\
\ee

\n
The cross-power spectra ${\cal J}_l^{AB^2,A}(r_1,r_2)$ can be simplified in terms of the following expression:

\be
{\cal J}_l^{AB^2,A}(r_1,r_2) =  \frac{1}{2l+1} \sum_{LM}(-1)^M \sum_m \left \{ F_{L}(r_1,r_2) \alpha_{l_1}(r_1)\alpha_{l_2}(r_1)\alpha_{l_2}(r_2)\beta_l(r_2) \right \} \langle a_{l_1m_1}a_{l_2m_2}a_{l_3m_3}a_{lm} \rangle {\cal G}_{l_1l_2L}^{m_1m_2M}{\cal G}_{l_3lL}^{m_3mM}.
\ee

\n
Where the Gaunt integral describing the integral involving three spherical harmonics is defined as follows:

\be
{\cal G}_{l_1l_2l_3}^{m_1m_2m_3} = \sqrt{(2l_1+1)(2l_2+1)(2l_3+1)\over 4 \pi}
\left ( \begin{array}{ c c c }
     l_1 & l_2 & l_3 \\
     0 & 0 & 0
  \end{array} \right)\left ( \begin{array}{ c c c }
     l_1 & l_2 & l_3 \\
     m_1 & m_2 & m_3
  \end{array} \right).\\
\ee

\n
The second terms can be treated in an analogus way and takes the following form:

\be
{\cal L}_l^{AB^2,M}(r) = \frac{1}{2l+1} \sum_{LM} (-1)^M \sum_m \left \{ \alpha_{l_1}(r_1)\alpha_{l_2}(r_1)\alpha_{l_2}(r_2)\mu_l(r_2) \right \}\langle a_{l_1m_1}a_{l_2m_2}a_{l_3m_3}a_{lm} \rangle {\cal G}_{l_1l_2L}^{l_1l_2L}{\cal G}_{l_3lL}^{l_3lL} .
\ee

\n
Finally when combined these terms as in Eq.(\ref{eq:k31}), we recover the following expression:

\be
{\cal K}^{(3,1)}_l = {1 \over 2l+1} \sum_{l_1l_2l_3} \sum_L {1 \over {\cal C}_{l_1}{\cal C}_{l_2}{\cal C}_{l_3}{\cal C}_l}
 T_{l_3l}^{l_1l_2}[L]T^{l_1l_2}_{l_3l}[L] .
\ee

\n
The estimator ${\cal K}_l^{(3,1)}$ depends linearly both on $f_{\rm NL}^2$ and $g_{\rm NL}$.  In principle, we can use the estimate of $f_{\rm NL}$ from a bispectrum analysis as  
a prior, or we can use the estimators ${\cal S}_l^{(2,1)}$,  ${\cal K}_l^{(3,1)}$ and  ${\cal K}_l^{(3,1)}$
to put joint constraints on $f_{NL}$ and $g_{NL}$. Computational evaluation of either of the power spectra 
clearly will be more involved as a double integral corresponding to two radial directions needs to be evaluated.
Given the low signal-to-noise associated with these power spectra, binning will be essential.

\subsection{Estimator for $\myK_l^{(2,2)}$}

\n
In an analogous way the other power spectra associated with the trispectra can be optimised by the following construction.
We start by taking the harmonic transform of the product field $A(r,\oh)B(r,\oh)$ evaluated at the same 
line-of-sight distance $r$:

\be
A(r,\oh)B(r,\oh)|_{lm} = \int A(r)B(r)~Y_{lm}(\oh)~d\oh; \qtwo
B(r,\oh)M(r,\oh)|_{lm} = \int B(r)M(r)~Y_{lm}(\oh)~d\oh, 
\ee

\n
and contract it with its counterpart at a different distance. The corresponding power spectrum
(which is a function of these two line-of-sight distances $r_1$ and $r_2$) has a first term

\be
{\cal J}_l^{AB,AB}(r_1,r_2) = \frac{1}{2l+1}\sum_m \sum_L F_L(r_1,r_2) A(r_1,\oh)B(r_1,\oh)|_{lm}A(r_2,\oh)B(r_2,\oh)|_{lm}^*  
\ee

\n
Similarly, the second part of the contribution can be constructed by cross-correlating the product of 3D fields
$A(\oh,r_1)B(\oh,r_1)$ and $B(\oh,r_2)M(\oh,r_2)$ evaluated at different $r_1$ and $r_2$,
with corresponding weighting function $F_L(r_1,r_2)$:

\be
{\cal L}_l^{AB,BM}(r) = \frac{1}{2l+1}\sum_m A(r,\oh)B(r,\oh)|_{lm}B(r,\oh)M(r,\oh)|_{lm}^* 
\ee

\n
Finally, the estimator is constructed as:

\be
{\cal K}_l^{(2,2)} = 4f_{nl}^2 \int r_1^2 dr_1 \int r_2^2 dr_2 {\cal J}_l^{AB,AB}(r_1,r_2) + 2g_{nl} \int r^2 dr {\cal L}_l^{AB,BM}(r) 
\label{eq:k22}
\ee

\n
To see they do correspond to an optimum estimator we use the harmonic expansions and follow the same procedure outlined before:

\ben
&& {\cal L}_l^{AB,BM}(r)  =  \frac{1}{2l+1}  \sum_m (-1)^m \sum_{l_im_i} \left \{ \alpha_{l_1}(r)\alpha_{l_2}(r)\beta_{l_3}(r)
\mu_{l_4}(r) \right \} \langle a_{l_1m_1}a_{l_2m_2}a_{l_3m_3}a_{l_4m_4} \rangle {\cal G}_{l_1l_2l}^{m_1m_2m}{\cal G}_{l_3l_4l}^{m_3m_4m} .\\
&& {\cal J}_l^{AB,AB}(r_1,r_2)  =  \frac{1}{2l+1} \sum_{LM} (-1)^M \sum_m \left \{ F_L(r_1,r_2)\alpha_{l_1}(r_1)\beta_{l_2}(r_1)\alpha_{l_2}(r_2)\beta_l(r_2) \right \} \langle a_{l_1m_1}a_{l_2m_2}a_{l_3m_3}a_{lm} \rangle {\cal G}_{l_1l_2L}^{m_1m_2M}{\cal G}_{l_3l_4L}^{m_3m_4M} .
\een

\n
Here we notice that ${\cal J}_l^{AB,AB}(r_1,r_2)$ is invariant under exchange of $r_1$ and $r_2$ but  ${\cal J}_l^{AB,BM}(r_1,r_2)$ is not.
Finally, joining the various contributions to construct the final estimator, as given in Eq.(\ref{eq:k22}), which involves a line-of-sight integration:

\be
{\cal K}_l^{(2,2)} = {1 \over 2l+1} \sum_{l_im_i} {1 \over \myC_{l1}\myC_{l_2}\myC_{l_3}\myC_{l_4}} { T^{l_1l_2}_{l_3l_4}(l) 
T^{l_1l_2}_{l_3l_4}(l) }.
\ee

\n
The prefactors associated with $f_{\rm NL}^2$ and $g_{\rm NL}$ are different in the linear combinations  ${\cal K}_l^{(2,2)}$ and ${\cal K}_l^{(3,1)}$, and hence even without using information
from third-order we can estimate both from fourth order alone.

\begin{figure}
\begin{center}
{\epsfxsize=9. cm \epsfysize=5. cm {\epsfbox[30 428 590 727]{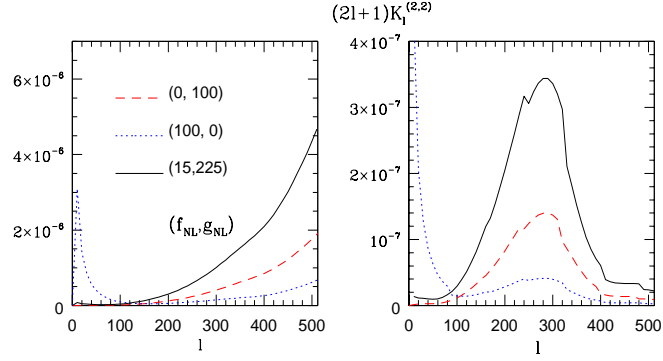}}}
\end{center}
\caption{The trispectrum-related power spectrum ${\cal K}^{(2,2)}_l$  is plotted as function
of angular scale $l$. The left panel is for a noise free ideal all-sky experiment. The right panel correspond to WMAP-V band noise.
Various curves correspond to a specific parameter combination of $(f_{NL},g_{NL})$ as indicated (see text for details).}
\label{fig:K22}
\end{figure}

\begin{figure}
\begin{center}
{\epsfxsize=9. cm \epsfysize=5. cm {\epsfbox[30 428 590 727]{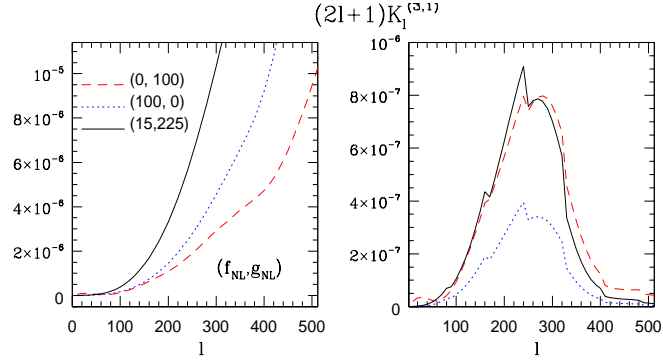}}}
\end{center}
\caption{Same as \ref{fig:K22} but for the trispectrum-related power spectrum ${\cal K}^{(3,1)}_l$ plotted as function of angular scale $l$
for ideal (left panel) and WMAP V-band noise charecteristics.}
\label{fig:K31}
\end{figure}

Figs. \ref{fig:K22} and \ref{fig:K31} shows the primordial $\myK_l^{(2,2)}$ and  $\myK_l^{(3,2)}$ for the WMAP5 best-fit cosmological parameters \citep{Dunkley}, integrated over a range of 500 Mpc around recombination, and shown as a function of harmonic number $l$. The computations of $\myK_l$ typically scales as $l_{max}^3$ with resolution which makes it quite
expensive for high resolution studies. The computations are largely dominated by computations of 3J symbols.
For a given configuration number of non-zero 3J symbols for which the computations are required roughly scales as 
$l_{max}^3$ which explains the scaling. For more details about noise and beam see \cite{Smidt09}

The unbiased version of $\myK_l^{(2,2)}$ has been used in the context of study of lensing effects on CMB maps \citep{CooKes03}.
While cross-correlational analysis can be helpful for detection of lensing on CMB maps for internal detection 
involving only CMB maps an analysis at the trispectrum level is necessary.

\subsection{Linking various Estimators}

\n
The one-point estimators can be expressed as a sum over all $l$s of the estimators $\myK_l^{(3,1)}$ or  $\myK_l^{(2,2)}$.
Therefore the corresponding optimised one-point estimators $\myK^{(4)}$ can be written as:

\be
\myK^{(4)} = \sum_{l_iL} { T_{l_1l_2}^{l_3l_4}(L)  \hat T_{l_1l_2}^{l_3l_4}(L) \over C_{l_1}C_{l_2}C_{l_3}C_{l_4}}.
\ee

\n
Here $\hat T_{l_1l_2}^{l_3l_4}(L)$ denotes the direct estimator from the harmonic transforms and  
$T_{l_1l_2}^{l_3l_4}(L)$ act as weights which are constructed using the optimisation function 
discussed above. Similarly, at the two-point level the corresponding power spectra 
can be written as:

\be
 {\myK}_l^{(2,2)} = \sum_{l_i} 
 { T_{l_1l_2}^{l_3l_4}(l) \hat T^{l_1l_2}_{l_3l_4}(l) \over C_{l_1}C_{l_2}C_{l_3}C_{l_4}}; \qtwo
{\myK}_l^{(3,1)} = \sum_{l_iL} { T^{l_1l_2}_{l_3l}(L) \hat T^{l_1l_2}_{l_3l}(L) \over C_{l_1}C_{l_2}C_{l_3}C_{l}}.
\ee

\n
Each contribution from a specific configuration formed by various values of $l_i$ and $L$ is weighted by
the corresponding $C_l$ to make the estimator optimal. For computation of the variances the fields are 
however considered as Gaussian, which should be a good  approximation as the fields are expected to be close to gaussian. 
Note that the $C_ls$ here take contributions from both the signal and the noise terms
i.e. $C_l = C_l^S + C_l^N$, where $C_l^S$ is just the theoretical expectation for primordial perturbations.

\n
Both ${\cal K}_l^{(2,2)}$ and ${\cal K}_l^{(3,1)}$ involve both parameters $f_{nl}$ and
$g_{nl}$ and can be used for a joint analysis, along with ${\cal S}_l^{(2,1)}$ to put constraints
from realistic data. It is easy to check that the following relationship holds:

\be
\myK^{(4)} = \sum_l(2l+1)K_l^{(2,2)} = \sum_l(2l+1)K_l^{(3,1)}.
\ee

\n
These results therefore generalises the results obtained in \citep{MuHe09} in the context of
bispectral analysis.

\subsection{Subtracting the Gaussian or disconnected Contributions}

In estimating the trispectrum we need to subtract out the disconnected or Gaussian 
parts \citep{Hu00,huOka02}.  To do this we follow
the same procedure but replacing the simulated non-Gaussian maps with their Gaussian
counterparts. Both maps should be constructed to have the same power spectrum, and identical
noise.  If the noise is Gaussian it will only contribute only to
the disconnected part. For construction of an estimator
which is noise-subtracted we need to follow the same procedure above by constructing Gaussian
maps $A^G(r,\oh), B^G(\oh,r)$ etc. 

\ben
&& A^G(\oh,r) = \sum_{lm} A^G_{lm}Y_{lm}(\oh); \qquad B^G(\oh,r) = \sum_{lm} B^G_{lm}Y_{lm}(\oh); \qquad  M^G(\oh,r) = \sum_{lm} M^G_{lm}Y_{lm}(\oh); \nn \\
&& A^G_{lm} =  {\alpha_{lm} \over C_l} a^G_{lm}; ~~ B^G_{lm} =  {\beta_{lm} \over C_l} a^G_{lm}; ~~ M^G_{lm} =  {\mu_{lm} \over C_l} a^G_{lm}. 
\een

\n
We illustrate this using ${\cal K}^{(2,2)}_l$; the analysis
is very similar for the other estimator. We start by replacing the quantities
${\cal I}_l^{AB,AB}(r_1,r_2)$ and  ${\cal L}_l^{AB,AB}(r_1,r_2)$ by their Gaussian 
counterparts, ${\cal P}_l^{A^GB^G,A^GB^G}(r_1,r_2)$ and
${\cal R}_l^{A^GB^G,B^GM^G}(r)$:

\ben
&& {\cal P}_l^{A^GB^G,A^GB^G}(r_1,r_2) = \frac{1}{2l+1}\sum_m \sum_L F_L(r_1,r_2) A^G(r_1,\oh)B^G(r_1,\oh)|_{lm}A^G(r_2,\oh)B^G(r_2,\oh)|_{lm}^* \nn \\
&&=  \frac{1}{2l+1} \sum_{LM}(-1)^M \sum_m \left \{ F_L(r_1,r_2)\alpha_{l_1}(r_1)\beta_{l_2}(r_1)\alpha_{l_2}(r_2)\beta_l(r_2) \right \} \langle a^G_{l_1m_1}a^G_{l_2m_2}a^G_{l_3m_3}a^G_{lm} \rangle {\cal G}_{l_1l_2L}^{m_1m_2M}{\cal G}_{l_3l_4L}^{m_3m_4M} .
\een

\n
These quantities are then used for the construction of an estimator, which in practice 
aims to estimate the Gaussian contributions to the total trispectra. 

\ben
&& {\cal R}_l^{A^GB^G,B^GM^G}(r) = \frac{1}{2l+1}\sum_m A^G(r,\oh)B^G(r,\oh)|_{lm}B^G(r,\oh)M^G(r,\oh)|_{lm}^* \nn \\
&& =  \frac{1}{2l+1}  \sum_m (-1)^m \sum_{l_im_i} \left \{ \alpha_{l_1}(r)\alpha_{l_2}(r)\beta_{l_3}(r)
\mu_{l_4}(r) \right \} \langle a^G_{l_1m_1}a^G_{l_2m_2}a^G_{l_3m_3}a^G_{l_4m_4} \rangle {\cal G}_{l_1l_2l}^{m_1m_2m}{\cal G}_{l_3l_4l}^{m_3m_4m}.
\een

\n
As before we combine these quantities to form the weighted Gaussian part of the contribution 
to the trispectrum:

\ben
{\cal G}_l^{(2,2)} = 4 f^2_{nl} \int r_1^2 dr_1 \int r_2^2 dr_2 {\cal P}(r_1,r_2) + 2g_{NL}\int r^2dr {\cal
R}(r) = {1 \over 2l+1} \sum_{l_i} { 1 \over C_{l_1}C_{l_2}C_{l_3}C_{l_4}}
 \hat G^{l_1l_2}_{l_3l_4}(l).
\een

\n
Subtracting the Gaussian contribution from the total estimator we can write:

\ben
{\cal K}_l^{(2,2)} ={\cal K}_l^{(2,2)} - {\cal G}_l^{(2,2)} = {1 \over 2l+1} \sum_{l_i} { 1 \over C_{l_1}C_{l_2}C_{l_3}C_{l_4}}
T^{l_1l_2}_{l_3l_4}(l)\left \{ \hat T^{l_1l_2}_{l_3l_4}(l) - \hat G^{l_1l_2}_{l_3l_4}(l)\right \}.
\een

\n
A very similar calculation for the other fourth-order estimator ${\cal G}_l^{3,1}$ provides
an identical result. After subtraction of the Gaussian contribution we can write:

\ben
{\cal K}_l^{(3,1)} = {\cal K}_l^{(3,1)} - {\cal G}_l^{(3,1)} = {1 \over 2l+1} \sum_{l_i}\sum_L { 1 \over C_{l_1}C_{l_2}C_{l_3}C_{l}}
T^{l_1l_2}_{l_3l_4}(l)\left \{ \hat T^{l_1l_2}_{l_3l}(L) - \hat G^{l_1l_2}_{l_3l}(L)\right \}.
\een

\n
The corresponding one-point collapsed estimator has the following form, and the relationship
between the various estimators remains unchanged.

\ben
{\cal K}_l^{(4)} = \sum_{l_i}\sum_L { 1 \over C_{l_1}C_{l_2}C_{l_3}C_{l_4}}
T^{l_1l_2}_{l_3l_4}(l)\left \{ \hat T^{l_1l_2}_{l_3l_4}(L) - \hat G^{l_1l_2}_{l_3l_4}(L)\right \}.
\een

\n
In the next section we consider partial sky coverage and the resulting corrections to the
Gaussian and non-Gaussian estimators.

\section{Partial Sky Coverage and Optimised Estimators}

\n
The terms that are needed for correcting the effect of finite sky coverage and 
inhomogeneous noise are listed below. These
corrections are incorporated by using Monte-carlo simulations of noise realisations \citep{Crem03,BaZa04,BCP04,Babich,Crem06,Crem07a,Crem07b}.
Whereas in case of bispectral analysis it is just the linear terms which are
needed for corrections, for trispectral analysis there are both linear and quadratic terms. 

We will treat the general case in later subsections, but first we present results for the simpler case of homogeneous noise and for high wavenumbers where the 
density of uncorrelated states is modified by inclusion of the fraction of sky observed, $f_{sky}$.

\subsection{Corrective terms for the Estimator~${\cal K}_l^{3,1}$ in the absence of spherical symmetry}

\ben
&&\hat {\cal J}_l^{A_1B_1B_2,A_2} = {1 \over f_{sky}} \left [ \tilde { \cal J}_l^{A_1B_1B_2,A_2} - {\cal I}^{\rm Lin}_l - 
{\cal I}^{\rm Quad}_l \right ] \\
&& { \cal I}^{\rm Lin}_l = {1 \over f_{sky}} \Big [
{\cal J}_l^{\langle A_1B_1 \rangle B_2, A_2}
+ {\cal J}_l^{\langle A_1B_2 \rangle B_1, A_2} 
+ {\cal J}_l^{\langle B_1B_2 \rangle A_1, A_2} 
+ {\cal J}_l^{ A_1B_1 \langle B_2, A_2 \rangle }
+ {\cal J}_l^{ A_1B_2 \langle B_1, A_2 \rangle} 
+ {\cal J}_l^{ B_1B_2 \langle A_1, A_2 \rangle} 
\Big ] \\ 
&& { \cal I}^{\rm Quad}_l= {1 \over f_{sky}} \Big [
{\cal J}_l^{\langle A_1B_1B_2 \rangle, A_2} 
+{\cal J}_l^{A_1 \langle B_1B_2, A_2 \rangle} 
+ {\cal J}_l^{B_1 \langle A_1B_2, A_2\rangle} 
+ {\cal J}_l^{B_2 \langle A_1B_1, A_2 \rangle} 
\Big ].
\een

\n
The expressions are similar for the other terms that depend on only one radial distance:

\ben
&&\hat {\cal L}_l^{ABB,M} = {1 \over f_{sky}} \left [ \tilde { \cal L}_l^{ABB,M} - {\cal I}^{Lin}_l - 
{\cal I}^{\rm Quad}_l \right ] \\
&& { \cal I}^{\rm Lin}_l= {1 \over f_{sky}} \Big [
{\cal L}_l^{A \langle BB, M \rangle} +  2{\cal L}_l^{B \langle A, M \rangle}
+2{\cal L}_l^{B \langle AB \rangle, M} \Big] \\
&& { \cal I}^{\rm Quad}_l= {1 \over f_{sky}} \Big [
{\cal L}_l^{A \langle BB \rangle, M} 
+2{\cal L}_l^{B \langle AB \rangle, M} \Big].
\een

\n
To simplify the presentation we have used the symbol $A(r_1,\oh)= A_1; A(r,\oh)=A$ and so on. Essentially we can see
that there are terms which are linear in the input harmonics and terms which are quadratic in the input
harmonics. The terms which are linear are also proportional to the bispectrum of the remaining 3D fields
which are being averaged. On the other hand the prefactors for quadratic terms are 
3D correlation functions of the remaining two fields. Finally putting all of these expressions we can write:

\be
\tilde {\cal K}_l^{(3,1)} = 4f_{\rm NL} \int r_1^2 dr_1 \int r_2 dr_2 \tilde {\cal J}_l^{AB^2,A}(r_1,r_2)  +
\int r^2 dr \tilde {\cal L}_l^{AB^2 M}(r). \\
\ee

\n
From a computational point of view clearly the overlap integral $F_{L}(r_1,r_2)$
will be expensive and may determine to what resolution ultimately these direct techniques can be
implemented. Use of these techniques directly involving Monte-Carlo numerical simulations will be
dealt with in a separate paper (Smidt et al. in preparation). To what extent the linear and quadratic
terms are important in each of these contributions can only be decided by testing against simulation.

\subsection{Corrective terms for the Estimator~ ${\cal K}_l^{2,2}$ in the absence of Spherical symmetry}

\n
The unbiased estimator for the other estimator can be constructed in a similar manner. As before there
are terms which are quadratic in input harmonics with a prefactor proportional to terms involving 
cross-correlation or variance of various combinations of 3D fields and there will be linear terms
(linear in input harmonics) with a prefactor proportional to bispectrum associated with various
3D fields.

\ben
&& \hat {\cal J}_l^{A_1B_1,A_2B_2} = {1\over f_{sky}} \Big [\tilde {\cal J}_l^{A_1B_1,A_2B_2} - I_l^{\rm Lin} - I_l^{\rm Quad}  \Big ]  \\
&&  {\cal I}^{\rm Lin} =  {1\over f_{sky}}\Big [ {\cal J}_l^{A_1B_1,\langle A_2B_2 \rangle} +{\cal J}_l^{\langle A_1B_1 \rangle,A_2B_2}+ 
{\cal J}_l^{A_1\langle B_1,B_2 \rangle A_2} + {\cal J}_l^{B_2 \langle A_2,B_2 \rangle A_2} +
{\cal J}_l^{B_2 \langle A_2,A_2 \rangle B_2} + {\cal J}_l^{A_2 \langle B_2,A_2 \rangle B_2} \Big ] \\
&& {\cal I}^{\rm Quad} = {1 \over f_{sky}}\Big [ {\cal J}_l^{A_1 \langle B_1, A_2B_2 \rangle} +{\cal J}_l^{B_1 \langle A_1, A_2B_2 \rangle} 
+ {\cal J}_l^{\langle A_1B_1, A_2 \rangle B_2} +{\cal J}_l^{\langle A_2B_2, B_2 \rangle A_2}  \Big ] .
\een

\n
The terms such as  ${\cal K}_l^{AB,BM}(r_1,r_2)$ can be constructed in a very similar way. We display
the term  ${\cal K}_l^{AB^2,A}(r_1,r_2)$ with all its corrections included. 

\ben
&& \hat {\cal L}_l^{AB,BM} = {1\over f_{sky}} \Big [\tilde {\cal L}_l^{AB,BM} - I_l^{\rm Lin} - I_l^{\rm Quad}  \Big ]  \\
&& {I}_l^{\rm Quad} = {1 \over f_{sky}}\Big [ {\cal L}_l^{A \langle B^2\rangle,A} + 2 {\cal L}_l^{AB\langle B,A \rangle}
 +{\cal K}_l^{B^2\langle A,A \rangle} +
2{\cal K}_l^{B\langle AB,A \rangle} + {\cal K}_l^{A\langle B^2,A \rangle} 
\Big \} \Big ] \\
&& {I}_l^{\rm Lin} = {1 \over f_{sky}}\Big [ {\cal J}_l^{A_1 \langle B_1, B_2M_2 \rangle} +{\cal J}_l^{B_1 \langle A_1, B_2M_2 \rangle} 
+ {\cal J}_l^{\langle A_1B_1, M_2 \rangle B_2} +{\cal J}_l^{\langle A_2B_2, B_2 \rangle M_2} \Big ] .
\een

The importance of the linear terms greatly depends on the target model being considered. For example, while linear terms
for bispectral analysis can greatly reduce the amount of scatter in the estimator for
{\it local} non-Gaussianity, the linear term is less important in modelling the {\it equilateral}  model. 
In any case the use of such Monte-Carlo (MC) maps is known to reduce the scatter and can greatly simplify 
the estimation of non-Gaussianity.  This can be useful, as 
fully optimal analysis with inverse variance weighting, which treats mode-mode coupling completely, may only be possible on low-resolution maps. 

\subsection{Corrective terms for the Estimator~${\cal K}^{(4)}$ in the absence of spherical symmetry}

\n
The corrections to the optimised one-point fourth order cumulant can also be analysed in a very similar manner.
It is expected that higher-order cumulants will be more affected by partial sky coverage and loss of
spherical symmetry because of inhomogeneous noise.

\be
\hat {\cal K}^{(4)} = {1 \over f_{sky}}[\tilde {\cal K}^{(4)} -   I^{\rm Lin} - I^{\rm Quad}  \Big ].
\ee

\n
The linear and quadratic terms can be expressed in terms of MC averages. We list terms which involve 
the overlap integral and the one with single line-of-sight integration.

\ben
&& I^{\rm Lin} = 4f_{\rm NL} \int r_1^2 dr_1 \int r_2 dr_2 \sum_L F^{12}_L \left \{ A_1\la B_1A_2B_2\ra + B_1\la A_1A_2B_2 \ra + A_2\la A_1B_1A_2 \ra +
 B_2\la A_1B_1A_2\ra\right \} \nonumber \\
&& \qtwo \qtwo \qtwo + 2g_{\rm NL} \int r^2 dr \left \{ A\la B^2M\ra + 2B\la ABM \ra + M \la AB^2\ra \right \} \\
&& I^{\rm Quad} = 4f_{\rm NL} \int r_1^2 dr_1 \int r_2 dr_2 \sum_L F^{12}_L \left \{ A_1B_1 \la A_2B_2 \ra + A_2B_2 \la A_1B_1\ra
+ A_1B_2\la A_2B_1\ra + A_2B_1\la A_1B_2 \ra \right \} \rangle \nonumber \\
&& \qtwo \qtwo \qtwo + 2g_{\rm NL} \int r^2 dr \left \{ AM\la B^2\ra + B^2\la AM \ra + 2BM \la AB\ra \right \} .
\een

\n
We have introduced the shorthand notation $F_L^{12}= F_L(r_1,r_2)$. 

\begin{table*}
\caption{Various multispectra and associated power-spectra are tabulated along with their cosmological use}
\begin{center}
\begin{tabular}{|c |c|c| c}
\hline
Order & Cumulants \& Correlator & Power Spectra & Use \\
\hline
$2=(1+1)$&$\langle\delta^2T(\oh)\rangle$, $\langle\delta T(\oh)\delta T(\oh')\rangle$ & $C_l$ & Constraints on Cosmology ($\Omega,H_0. \dots$) \\
\hline
$3=(2+1)$& $\langle\delta^3T(\oh)\rangle$, $\langle\delta^2 T(\oh)\delta T(\oh')\rangle$ & $S^{(3)}$, ~$S_l^{(2,1)}$ & Inflationary Models $f_{\rm NL}$, Secondaries
 x Lensing $b_l$ \\
\hline
$4=(3+1),(2+2)$& $\langle\delta^4T(\oh)\rangle$, $\langle \delta^3 T(\oh)\delta(\oh')\rangle$,$\langle \delta^2 T(\oh)\delta^2 T(\oh')\rangle$ 
&$K^{(4)}$,~ $K_l^{(2,2)}$,~$K_l^{(3,1)}$ & $f_{\rm NL}, g_{\rm NL}$, KSZ-Lensing Sep.
Internal Lensing det.\\
\hline
\label{table:multispectra}
\end{tabular}
\end{center}
\end{table*}

\section{Realistic Survey Strategies: Exact Analysis }

In this section, we include the full optimisation, including the mode-mode coupling introduced by partial sky coverage and inhomogeneous noise, generalising results from the three-point level \citep{Babich,SmZa06,SmZaDo00,SmSeZa09}.   As before, we find that for the trispectrum, the addition of quadratic terms in addition to linera terms is needed.   The analysis presented here is completely generic and will not depend on details of
factorizability properties of the trispectra. For any specific form of the trispectrum 
the technique presented here can always provide optimal estimators. Importantly, this makes
it suitable also for the study of the trispectrum contribution from secondaries, and offers the possibility of determining whether any observed connected trispectrum is primordial or not.  Generalisation to multiple sources of trispectrum is straightforward, following \cite{Smidt09}

Trispectra from secondary anisotropies such as gravitational lensing are expected to dominate
the contribution from primary anisotropies \citep{CooKes03}. The estimator we develop here will be directly
applicable to data from various surveys, but the required direct inversion of the covariance matrix in the harmonic domain 
may not be computationally feasible 
in near future. Nevertheless an exact analysis may still be beneficial for 
low-resolution degraded maps where the primary anisotropy dominates. At higher resolution 
the exact analysis will reduce to the one discussed in previous sections.
In addition it may be possible at least to certain resolution to bypass the exact 
inverse variance weighting by introducing a conjugate gradient techniques.

The general theory for optimal estimation from data was developed by \cite{Babich} for the
analysis of the bispectrum, and was later implemented by \cite{SmZa06}. For arbitrary  sky coverage and inhomogeneous noise the estimator will be fourth order in input harmonics, and involves matched filtering to maximise the response of the estimator when the estimated trispectrum matches 
theoretical expectation. 
We present results for both one-point cumulants and two-point cumulant correlators or power spectra associated with trispectra.
The estimators presented here, $E_l^{(3,1)}$ and $E_l^{(2,2)}$ are generalisations of estimators $C_l^{(3,1)}$, $C_l^{(2,2)}$ and
$\myK^{(3,1)}$ and $\myK^{(2,2)}$ presented in previous sections.

\subsection{One-point Estimators}

\n
We will use inverse variance weighting harmonics recovered from the sky. The covariance matrix, expressed
in the harmonics domain, $C^{-1}_{LM,lm}$ when used to filter out modes recovered directly from sky are $a_{lm}$ 
are denoted as $A_{LM}$. We use these harmonics to construct optimal estimators. For all-sky coverage
and homogeneous noise, we can we recover $A_{lm} = a_{lm}/C_l$, with $C_l$s including signal and noise. 
We start by keeping in mind that the trispectrum can be expressed in terms of the
harmonic transforms $a_{lm}$ as follows:

\be
T^{l_al_b}_{l_cl_d}(L) = (2l+1) \sum_{m_i} \sum_M (-1)^M 
\left ( \begin{array}{ c c c }
     l_a & l_b & L \\
     m_a & m_b & M
  \end{array} \right)
\left ( \begin{array}{ c c c }
     l_c & l_d & L \\
     m_c & m_d & -M
  \end{array} \right) 
a_{l_am_a} \dots a_{l_dm_d} \qtwo (i \in {a,b,c,d}).
\ee

\n
Based on this expression we can devise a one-point estimator. In our following discussion, the relevant harmonics can be 
based on partial sky coverage. 

\be
Q^{(4)}[a] = {1 \over 4!} \sum_{LM} (-1)^M \sum_{l_im_i} \Delta(l_i;L) ~ T^{l_1l_2}_{l_3l_4}(L) 
 \left ( \begin{array}{ c c c }
     l_a & l_b & L \\
     m_a & m_b & M
  \end{array} \right)
\left ( \begin{array}{ c c c }
     l_c & l_d & L \\
     m_c & m_d & -M
  \end{array} \right)
a_{l_1m_1} \dots a_{l_4m_4}
\ee

\n
The term $\Delta(l_i,L)$ is introduced here to avoid contributions from Gaussian or disconnected 
contributions. $\Delta(l_i,L)$ vanishes if amy pair of $l_i$s becomes equal or $L=0$ which effectively
reduces the trispectra to a product of two power spectra (i.e. disconnected Gaussian pieces).

We will also need the first-order and second-order derivative with respect to the input harmonics.
The linear terms are proportional to the first derivatives and the quadratic terms are proportional
to second derivatives of the function $Q[a]$, which is quartic in input harmonics.

\be
\partial_{lm} Q^{(4)}[a] =  {1 \over 3!} \sum_{LM} (-1)^M  \sum_{l_im_i} \Delta(l_i;L) ~  T^{l_1l_2}_{l_3l_4}(L)
\left ( \begin{array}{ c c c }
     l & l_a & L \\
     m & m_a & M
  \end{array} \right)
\left ( \begin{array}{ c c c }
     l_b & l_c & L \\
     m_b & m_c & -M
  \end{array} \right)
a_{l_am_b}a_{l_am_b}a_{l_cm_c}.
\ee

\n
The first-order derivative term $\partial_{lm} Q^{(4)}[a]$ is cubic in input maps and the second order derivative
is quadratic in input maps (harmonics). However unlike the estimator itself  $Q^{(4)}[a]$, which is simply a number,
these objects represent maps constructed from harmonics of the observed maps.

\be
\partial_{lm}\partial_{l'm'} Q^{(4)}[a] =  {1 \over 2!} \sum_{LM} (-1)^M \sum_{l_im_i} \Delta(l_i;L) ~  T^{l_1l_2}_{l_3l_4}(L)
\left ( \begin{array}{ c c c }
     l & l_a & L \\
     m & m_a & M
  \end{array} \right)
\left ( \begin{array}{ c c c }
     l & l_b & L \\
     m & m_b & -M
  \end{array} \right)
a_{l_am_a}a_{l_bm_b}.
\ee

\n
The optimal estimator for the one-point cumulant can now be written as follows. This is optimal
in the presence of partial sky coverage and most general inhomogeneous noise:

\be
E^{(4)}[a] =  {1 \over N}  \left \{ Q^{(4)}[C^{-1}a] -[C^{-1}a]_{lm} \langle \partial_{lm}Q^{(4)}[A] \rangle -[C^{-1}a]_{lm}[C^{-1}a]_{l'm'} \langle 
\partial_{lm}\partial_{l'm'}Q^{(4)}[C^{-1}a] \rangle \right \}.
\ee

\n
The terms which are subtracted out are linear and quadratic in input harmonics. The linear term 
is similar to the one which is used for bispectrum estimation, whereas the quadratic terms correspond to
the disconnected contributions and will vanish identically as we have designed our estimators in such
a way that it will not take any contribution from disconnected Gaussian terms. The Fisher matrix
reduces to a number which we have used for normalisation.

\ben
F= {1 \over N} = &&{1 \over 4!} \sum_{LM} \sum_{L'M'} \sum_{(all~lm)} \sum_{(all~l'm')} (-1)^M (-1)^{M'}~\Delta(l_i;L)\Delta(l_i';L')~ T^{l_al_b}_{l_cl_c}(L) T^{l_{a'}l_{b'}}_{l_{c'}l_{d'}}(L)
\left ( \begin{array}{ c c c }
     l_a & l_b & L \\
     m_a & m_b & M
  \end{array} \right)
\left ( \begin{array}{ c c c }
     l_c & l_d & L \\
     m_c & m_d & -M
  \end{array} \right) \nonumber \\
&& \qtwo \times  \left ( \begin{array}{ c c c }
     l_a' & l_b' & L' \\
     m_a' & m_b' & M'
  \end{array} \right)
\left ( \begin{array}{ c c c }
     l_c' & l_d' & L' \\
     m_c' & m_d' & -M'
  \end{array} \right) C^{-1}_{l_am_a,l_a'm_a'}\dots C^{-1}_{l_dm_d,l_d'm_d'}.
\een

\n
The ensemble average of this one-point estimator will be a linear combination of parameters $f_{\rm NL}^2$ and $g_{\rm NL}$.
Estimators constructed at the level of three-point cumulants \citep{SmZa06,MuHe09} can be used jointly
with this estimator to put independent constraints separately on  $f_{\rm NL}$ and $g_{\rm NL}$.
As discussed before, while one-point estimator has the advantage of higher signal-to-noise, such estimators are
not immune to contributions from an unknown component which may not have cosmological origin, such as inadequate foreground separation. 
The study of these power spectra associated with bispectra or trispectra
can be useful in this direction. Note that these direct estimators are computationally expensive due to the inversion and multiplication of large matrices, but can be implemented in low-resolution studies where primordial signals may be less contaminated by
foreground contributions or secondaries.

\subsection{Two-point Estimators: Power Spectra associated with Trispectra}

Generalising the above expressions for the case of the power spectrum associated with trispectra, we recover the two power spectra 
we have discussed in previous sections. The information content in these power spectra are optimal, and when summed for
over $L$ we can recover the results of one-point estimators. 

\be
Q^{(2,2)}_L[a] = {1 \over 4!} \sum_{M} (-1)^M \sum_{l_im_i} T^{l_al_b}_{l_cl_d}(L)
 \left ( \begin{array}{ c c c }
     l_a & l_b & L \\
     m_a & m_b & M
  \end{array} \right)
\left ( \begin{array}{ c c c }
     l_c & l_d & L \\
     m_c & m_d & -M
  \end{array} \right)
a_{l_am_a}\dots a_{l_dm_d}.
\ee

\n
The derivatives at first order and second order are as series of maps (for each $L$) constructed from the harmonics of the observed sky.
These are used in the construction of linear and quadratic terms. 

\be
\partial_{lm} Q^{(2,2)}_L[a] = {1 \over 3!} \sum_{T} (-1)^M \sum_{l_im_i}\Delta(l_i;L) T^{l_al_b}_{l_cl_d}(L)
 \left ( \begin{array}{ c c c }
     l & l_a & L \\
     m & m_b & M
  \end{array} \right)
\left ( \begin{array}{ c c c }
     l_b & l_c & L \\
     m_b & m_c & -M
  \end{array} \right)
a_{l_am_a}\dots a_{l_cm_c}.
\ee

\n
We can construct the other estimator in a similar manner. To start with we define the function $Q^{(3,1)}_L[a]$
and construct its first and second derivatives. These are eventually used for construction of the
estimator $E^{(3,1)}_L[a]$. As we have seen, both of these estimators can be collapsed to a one-point estimator
$Q^{(4)}[a]$. As before, the variable $a$ here denotes input harmonics $a_{lm}$ recovered from the noisy observed sky.

\be
Q^{(3,1)}_L[a] = {1 \over 3!} \sum_{M} \sum_{ST} (-1)^T  a_{LM} \sum_{l_im_i} \Delta(l_i,L;T) T^{Ll_b}_{l_cl_d}(T)
 \left ( \begin{array}{ c c c }
     L & l_b & S \\
     M & m_b & T
  \end{array} \right)
\left ( \begin{array}{ c c c }
     l_c & l_d & S \\
     m_c & m_d & -T
  \end{array} \right)
a_{l_bm_b}\dots a_{l_dm_d}.
\ee

\n
The derivative term will have two contributing terms corresponding to the derivative w.r.t. the free index $\{LM\}$
and the terms where indices are summed over e.g. $\{lm\}$, which is very similar to the results for the bispectrum
analysis with the estimator $Q^{(2,1)}_L[a]$. One major difference that needs to be taken into account is 
the subtraction of the Gaussian contribution. The function $\Delta(l_i,L)$ takes into account of this subtraction.

\ben
\partial_{lm} Q^{(3,1)}_L[a] = && \sum_{ST}  \sum_{l_im_i} \Delta(l_i,L;T) T^{Ll_2}_{l_3l_4}(T)
 \left ( \begin{array}{ c c c }
     l & l_b & S \\
     m & m_b & T
  \end{array} \right)
\left ( \begin{array}{ c c c }
     l_c & l_d & S \\
     m_c & m_d & -T
  \end{array} \right)
a_{l_bm_b}\dots a_{l_dm_d} \\
&& +3 \sum_M \sum_{l_im_i} a_{LM} \sum_T
 \left ( \begin{array}{ c c c }
     L & l & S \\
     M & m & T
  \end{array} \right)
\left ( \begin{array}{ c c c }
     l_b & l_c & S \\
     m_b & m_c & -T
  \end{array} \right)
a_{l_bm_b}a_{l_cm_c}.
\een

\n
Using these derivatives we can construct the estimators $E_L^{(3,1)}$ and $E_L^{(2,2)}$:

\ben
E_L^{(3,1)} = N_{LL'}^{-1} \left \{ Q^{(3,1)}_{L'}[C^{-1}a] -[C^{-1}a]_{lm} \langle \partial_{lm}Q^{(3,1)}_{L'}[C^{-1}a] \rangle \right \} \\
E_L^{(2,2)} = N_{LL'}^{-1} \left \{ Q^{(2,2)}_{L'}[C^{-1}a] -[C^{-1}a]_{lm} \langle \partial_{lm}Q^{(2,2)}_{L'}[C^{-1}a] \rangle \right \},
\een

\n
where summation over $L'$ is implied.  The quadratic terms will vanish, as they contribute only to the disconnected part. The normalisation constants are the Fisher matrix elements $F_{LL'}$  which can be expressed in terms of the target trispectra 
$T^{l_1l_2}_{l_3l_4}(L)$ and inverse covariance matrices $C^{-1}$ used for the construction of these estimators. 
The Fisher matrix for the estimator $E_L^{(3,1)}$, i.e. $F_{LL'}^{(3,1)}$ can be expressed as:

\ben
&& [N^{-1}]_{LL'} = F_{LL'}^{(2,2)} =  \left ( {1 \over 4!} \right )^2 \sum_{ST,S'T'}\sum_{(all~lm,l'm')}(-1)^M (-1)^{M'} [T^{l_al_b}_{l_cl_c}(L)] [T^{l_{a'}l_{b'}}_{l_{c'}l_{d'}}(L')] \nonumber \\ 
&& \qtwo \Delta(l_il;L)\Delta(l_i';L') \times \left ( \begin{array}{ c c c }
     l_a & l_b & S\\
     m_a & m_b & T
  \end{array} \right)
\left ( \begin{array}{ c c c }
     L & l_d & S \\
     M & m_d & -T
  \end{array} \right) 
\left ( \begin{array}{ c c c }
     l_a' & l_b' & S' \\
     m_a' & m_b' & T'
  \end{array} \right)
\left ( \begin{array}{ c c c }
     L' & l_d' & S' \\
     M' & m_d' & -T'
  \end{array} \right) \nonumber \\
&& \qtwo \times \Big \{ 6 C^{-1}_{LM,L'M'}C^{-1}_{l_am_a,l_a'm_a'}C^{-1}_{l_bm_b,l_b'm_b'}C^{-1}_{l_cm_c,l_c'm_c'}  
+ 18 C^{-1}_{LM,l_a'm_a'}C^{-1}_{l_am_a,L'M'}C^{-1}_{l_bm_b,l_b'm_b'}C^{-1}_{l_cm_c,l_c'm_c'} \Big \} .
\een

\n
Similarly for the other estimator $E_L^{(2,2)}$, the Fisher matrix $F_{LL'}^{(2,2)}$ can be written as a function of the associated trispectrum and the covariance matrix of various modes. For further simplification of these expressions we need to make simplifying assumptions for a specific type of trispectra, see \cite{MuHe09} for more details for such simplifications in the bispectrum.

\ben
[N^{-1}]_{LL'} = F_{LL'}^{(3,1)} = \left ( {1 \over 4!} \right )^2 \sum_{M} \sum_{M'} \sum_{(all~lm)} \sum_{(all~l'm')} (-1)^M (-1)^{M'} T^{l_al_b}_{l_cl_c}(L) T^{l_{a'}l_{b'}}_{l_{c'}l_{d'}}(L')
\left ( \begin{array}{ c c c }
     l_a & l_b & L \\
     m_a & m_b & M
  \end{array} \right)
\left ( \begin{array}{ c c c }
     L & l_d & L \\
     M & m_d & -M
  \end{array} \right) \nonumber \\
\times \left ( \begin{array}{ c c c }
     l_a' & l_b' & L' \\
     m_a' & m_b' & M'
  \end{array} \right)
\left ( \begin{array}{ c c c }
     L' & l_d' & L' \\
     M' & m_d' & -M'
  \end{array} \right) \Delta(l_il;L)\Delta(l_i';L')~C^{-1}_{l_am_a,l_a'm_a'}\dots C^{-1}_{l_dm_d,l_d'm_d'}.
\een

\n
Knowledge of sky coverage and the noise characteristics resulting from a specific scanning strategy etc. is needed
for modelling of  $C^{-1}_{l_dm_d,l_d'm_d'}$. We will discuss the impact of inaccurate modelling
of the covariance matrix in the next section. The direct summation we have used for the construction of the Fisher matrix may not
be feasible except for low resolution studies. However a hybrid method may be employed to combine the estimates
from low resolution maps using exact method with estimates from higher resolution using other faster but optimal
techniques described in previous section. In certain situations when the data is noise-dominated further
approximation can be made to simplify the implementation. A more detailed discussion will be presented elsewhere.

\subsection{Approximation to exact $C^{-1}$ weighting and non-optimal weighting}

If the covariance matrix is not accurately known due to the lack of exact beam or noise characteristics, or due to limitations on computer resources,
it can be approximated.  
An approximation $R$ of $C^{-1}$ then acts as a regularisation method.
The corresponding generic estimator can then be expressed as:

\be
\hat E_L^{Z}[a] = \sum_{L'} [F^{-1}]_{LL'} \left \{ Q_{L'}^Z[Ra]- [Ra]_{lm} \langle \partial_{lm}Q^Z_{L'}[Ra] \rangle \right \};
\qtwo Z \in \{(2,2),(3,1)\}.
\ee

\n
As before we have assumed sums over repeated indices and $\langle \cdot \rangle$ denotes Monte-Carlo (MC) averages.
As evident from the notations, the estimator above can be of type $E_L^{(3,1)}$ or $E_L^{(2,2)}$. For the collapsed case $E_L^{(4)}$ can also be handled in a very similar manner.  

\be
\hat E[a] =  { [F^{-1}]_{LL'}} \left \{ Q_L[Ra]- [Ra]_{lm} \langle \partial_{lm}Q_L[Ra] \rangle \right \}.
\ee
We will drop the superscript $Z$ for simplicity but any conclusion drawn below will be valid for both specific cases i.e.
$Z \in \{(2,2),(3,1)\}$. The normalisation constant which acts also as inverse of associated Fisher matrix $F_{LL'}$ can be written as:

\be
F_{LL'} = \langle({\hat E_L})(\hat E_{L'})\rangle  -\langle({\hat E_L})\rangle \langle (\hat E_{L'})\rangle 
 = {1 \over 4}\langle \partial_{lm}Q_L[Ra] \partial_{lm}Q_L[Ra] \rangle - {1 \over 4}\langle \partial_{lm}Q_L[Ra] \rangle \langle \partial_{lm}Q_{L'}[Ra] \rangle .
\ee

\n
The construction of $F_{LL'}$ is equivalent to the calculation presented for the case of $R=C^{-1}$. For
one-point estimator we similarly can write $F^R = \sum_{LL'} F_{LL'}^{R}$. The optimal weighting can be 
replaced by arbitrary weighting. As a spacial case we can also use no weighting at all
$R = I$. Although the estimator remains unbiased the scatter however increases as the estimator is no longer 
optimal. Use of arbitrary  weights makes the estimator equivalent to a PCL estimator.

\subsection{Joint Estimation of Multiple trispectra}

It may be of interest to estimate several trispectra jointly. In such scenarios it is indeed important
to construct a joint Fisher matrix which will 

\be
\hat E_L^{X}[a] = \sum_{XY} \sum_{LL'} [F^{-1}]^{XY}_{LL'} \hat E_{L'}^{Y}[a].
\ee

\n
The estimator $\hat E_L^{X}[a]$ is generic and it could be either $E^{(3,1)}$ or $E^{(2,2)}$.
Here $X$ and $Y$ corresponds to different trispectra of type $X$ and $Y$, these could be e.g. 
primordial trispectra from various inflationary scenarios. It is possible of course to do a joint estimation of 
primary and secondary trispectra. The off-diagonal blocks of the Fisher matrix will correspond to
cross-talks between various types of bispectra. Indeed principal component analysis or Generalised
eigenmode analysis can be useful in finding how many independent components of such trispectra can
be estimated from the data.

The cross terms in the Fisher matrix elements will be of following type:

\ben
&& F^{XY}_{LL'} =  \left ( {1 \over 4!} \right )^2 \sum_{ST,S'T'}\sum_{(all~lm,l'm')}(-1)^M (-1)^{M'} [T^{l_al_b}_{l_cl_c}(L)]^X [T^{l_{a'}l_{b'}}_{l_{c'}l_{d'}}(L')]^Y \nonumber \\ 
&& \qtwo \times \left ( \begin{array}{ c c c }
     l_a & l_b & S\\
     m_a & m_b & T
  \end{array} \right)
\left ( \begin{array}{ c c c }
     L & l_d & S \\
     M & m_d & -T
  \end{array} \right) 
\left ( \begin{array}{ c c c }
     l_a' & l_b' & S' \\
     m_a' & m_b' & T'
  \end{array} \right)
\left ( \begin{array}{ c c c }
     L' & l_d' & S' \\
     M' & m_d' & -T'
  \end{array} \right) \nonumber \\
&& \qtwo \times \Big \{ 6 C^{-1}_{LM,L'M'}C^{-1}_{l_am_a,l_a'm_a'}C^{-1}_{l_bm_b,l_b'm_b'}C^{-1}_{l_cm_c,l_c'm_c'}  
+ 18 C^{-1}_{LM,l_a'm_a'}C^{-1}_{l_am_a,L'M'}C^{-1}_{l_bm_b,l_b'm_b'}C^{-1}_{l_cm_c,l_c'm_c'} \Big \} .
\een

\n
The expression displayed above is valid only for $E^{(3,1)}$, exactly similar results holds for 
the other estimator $E^{(3,1)}$. For $X=Y$ we recover the results presented in previous section
for independent estimates. As before we recover the usual result for one-point estimator for
$Q^{4}$ from the Fisher matrix of $Q_L^{(3,1)}$ or $Q_L^{(2,2)}$, with corresponding estimator
modified accordingly.

\be
F^{XY} = \sum_{LL'} F^{XY}_{LL'}; \qtwo \hat E^{X}[a] = \sum_{XY} [F^{-1}]^{XY} \hat E^{Y}[a].
\ee

A joint estimation can provide clues to cross-contamination from different sources of trispectra.
It also provide information about the level of degeneracy involved in such estimates.

\section{Conclusions}

In the near future the all-sky Planck satellite will complete mapping the CMB sky in unprecedented detail, covering
a huge frequency range. The cosmological community will have the opportunity to use the resulting data to 
constrain available theoretical models. While the power spectrum provides the bulk of the information,
going beyond this level will lift degeneracies among various early universe scenarios which 
otherwise have near identical power spectra. The higher-order spectra are the harmonic transforms of
multi-point correlation functions, which contain information which can be difficult
to extract using conventional techniques. This is related to their complicated response to inhomogeneous
noise and sky coverage. A practical advance is to form collapsed two-point statistics, constructed from higher-order correlations, which can 
be extracted using conventional power-spectrum estimation methods. 

We have specifically studied and developed three different types of estimators which can be employed
to analyse these power spectra associated with higher-order statistics. The MASTER-based approach  \citep{Hiv} 
is typically employed to estimate pseudo-$C_l$s from the masked sky in the presence of noise; also see \cite{Efs1,Efs2}.
These are unbiased estimators but the associated variances and scatter can be estimated analytically
with very few simplifying assumptions. We extend these estimators to study higher-order correlation
functions. We develop estimators for $C_l^{(2,1)}$ for the skew-spectrum (3-point) as well as $C_l^{(3,1)}$ and $C_l^{(2,2)}$ which are power spectra of fields related to the trispectrum,
or kurt-spectrum (4-point). The removal of the Gaussian contribution is achieved by applying the 
same estimators to a set of Gaussian simulations with identical power spectra and subtracting.

As a next step we generalised the estimators employed by \cite{YKW,YaWa08,Yadav08} and others to study the kurt-spectrum. These methods are computationally expensive and can be implemented
using a Monte-Carlo pipeline which can generate 3D maps from the cut-sky harmonics using 
radial integrations of a target theoretical model. The Monte-Carlo generation of 3D maps is the most computationally expensive
part and dominates the calculation. The technique nevertheless has been used extensively,
as it remains highly parallelisable and is optimal in the presence of homogeneous noise and near
all-sky coverage. The corrective terms involves linear and quadratic terms for lack of spherical
symmetry due to inhomogeneous noise and partial sky coverage. These terms can be computed
using a Monte-Carlo chain. We also showed that the  radial integral involved at the three-point
analysis needs to extended to a double-integral for the trispectrum.
The speed of this analysis depends on how fast we can generate non-Gaussian maps. It is possible also
to use the same formalism to study cross-contamination from other sources such as 
point sources or other secondaries while also determining primordial non-Gaussianity. The
analysis also allows us to compute the overlap or degeneracy among various theoretical models
for primordial non-Gaussianity.

Finally we presented the analysis for the case of estimators which are completely optimal even in the
presence of inhomogeneous noise and arbitrary sky coverage \citep{SmZa06}. Extending previous work by \cite{MuHe09}
which concentrated only on the skew-spectrum, we showed how to generalise to the trispectrum. This involves finding a fast method
to construct and invert the covariance matrix $C_{lml'm'}$ in multipole space.
In most practical circumstances it is possible only to find an approximation to
the exact covariance matrix, and to cover this we present analysis for an approximate
matrix which can be used as instead of $C^{-1}$. This makes the method marginally suboptimal
but it remains unbiased. The four-point correlation function also takes contributions which
are purely Gaussian in nature. The subtraction of these contributions is again simplified by the
use of Gaussian Monte-Carlo maps with the same power spectrum. A Fisher analysis was presented 
for the construction of the error covariance matrix, allowing joint estimation of trispectra contributions from various sources, primaries
or secondaries. Such a joint estimation give us fundamental limits
on how many  sources of non-Gaussianity can be jointly 
estimated from a specific experimental set up. A more detailed Karhuenen-Lo\'eve eigenmode analysis will be
presented elsewhere.  The detection of non-primordial effects, such as weak lensing of the CMB, the kinetic SZ effect, the Ostriker-Vishniac effect and the thermal SZ effect can provide valuable additional cosmological information
\citep{RiqSper}.    The detection of CMB lensing at the level of the bispectrum 
needs external data sets which trace large-scale structures, but the trispectrum offers an internal detection, albeit with reduced S/N,  providing a valuable
consistency check. 

At the level of the bispectrum, primordial non-Gaussianity can for many models be described by a single parameter $f_{NL}$.
The two degenerate kurt-spectra (power spectra related to tri-spectrum) we have studied at the 4-point level require typically
two parameters such as $f_{NL}$ and $g_{NL}$. Use of the two power spectra will enable
us to put separate constraints on $f_{NL}$ and $g_{NL}$ without using information from lower-order
analysis of bispectrum, but they can all be used in combination. Clearly at even higher-order more parameters will be needed to
describe various parameters ($f_{NL}$, $g_{NL}$, $h_{NL}$, \dots) which will all be essential
in describing degenerate sets of power spectra associated with multispectra at a specific level.
Note we should keep in mind that higher-order spectra will be progressively more dominated by noise and
may not provide useful information beyond a certain point.

The power of the estimators we have constructed largely depends on finding a techniques to
simulate non-Gaussian maps with a specified bispectrum and trispectrum. 
We have discussed the possibility of using our technique to generate all-sky CMB maps 
with specified lower-order spectra, i.e. power spectrum, bispectra and trispectra.
These will generalise previous results by \cite{SmZa06} which can guarantee a specific
form at bispectrum level. 

While we have primarily focused on CMB studies, our estimators can also be useful in other areas 
e.g. studies involving 21cm studies, near all-sky redshift surveys and weak lensing
surveys. The estimators described here can be useful for testing theories for primordial 
and/or gravity-induced secondary trispectrum
using such diverse data sets. In the near future all-sky CMB experiments such as Planck can provide
maps covering a huge frequency range and near-all sky coverage. The estimators described here
can play a valuable role in analysing such maps.

\section{Acknowledgements}
\label{acknow}
The initial phase of this work was completed when DM was supported by a STFC rolling grant at the Royal Observatory,
Institute for Astronomy, Edinburgh. DM also acknowledges support from STFC rolling grant ST/G002231/1
at School of Physics and Astronomy at Cardiff University where this work was completed.
AC, PS, and JS acknowledge support from NSF AST-0645427. DM acknowledges
useful exchanges with Alexei Starobinsky, Patrick Valageas and Wayne Hu.

\bibliography{paper.bbl}

\appendix

\section{3j Symbols}

\ben
&& \sum_{l_3m_3} (2l_3+1) \left ( \begin{array}{ c c c }
     l_1 & l_2 & l_3 \\
     m_1 & m_2 & m_3
  \end{array} \right )
\left ( \begin{array}{ c c c }
     l_1 & l_2 & l \\
     m_1' & m_2' & m
  \end{array} \right ) = \delta_{m_1m_1'} \delta_{m_2m_2'} \\
&& \sum_{m_1m_2} \left ( \begin{array}{ c c c }
     l_1 & l_2 & l_3 \\
     m_1 & m_2 & m_3
  \end{array} \right)
\left ( \begin{array}{ c c c }
     l_1 & l_2 & l_3' \\
     m_1 & m_2 & m_3'
  \end{array} \right) = {\mathcal \delta_{l_3l_3'} \delta_{m_3m_3'} \over 2l_3 + 1} \\
&& (-1)^m\left ( \begin{array}{ c c c }
     l & l & l' \\
     m & -m & 0
  \end{array} \right ) = {(-1)^l \over \sqrt{(2l+1)}} \delta_{l'0}
\een

\end{document}